\documentclass[11pt, a4paper]{article}
\pdfoutput=1 
\usepackage[height=8.85in,width=6.75in]{geometry}
\usepackage{graphicx,rotating}     
\usepackage[bookmarksopen,colorlinks=true,linkcolor=teal,
citecolor=orangered,urlcolor=orangered,linktoc=all]{hyperref}

\usepackage{amsmath,amssymb}
\usepackage{slashed}
\usepackage{bm}
\usepackage{bbm}
\usepackage{cite}
\usepackage{soul}
\usepackage{comment}
\usepackage[utf8]{inputenc}
\usepackage{accents}
\usepackage[footnotesize]{caption}
\usepackage{cancel}
\usepackage{xstring}

\makeatletter
\@addtoreset{equation}{section}

\makeatother

\usepackage{xcolor}
\definecolor{teal}{rgb}{0,0.502,0.502}
\definecolor{orangered}{rgb}{1,0.27,0.}
\definecolor{redorange}{rgb}{1,0.325,0.286}
\definecolor{forestgreen}{rgb}{0.13,0.55,0.13}

\usepackage{tikz}
\usepackage{tikz-feynman}
\tikzfeynmanset{compat=1.0.0}
\pgfdeclarelayer{bg}    
\pgfsetlayers{bg,main}  

\newcommand{\brr}[1]{\left(#1\right)}

\newcommand{\hBf}[1]{\hat{{\bf #1}}}
\newcommand{\angf}[1]{\langle #1 \rangle}
\newcommand{\squf}[1]{[ #1 ]}

\newcommand{\IsBoldNumber}[1]{%
	\IfSubStr{1234}{#1}{\mathbf{#1}}{#1}%
}
\newcounter{i}
\newcommand{\bn}[1]{%
	\StrLen{#1}[\len]%
	\edef\result{}%
	\setcounter{i}{1}%
	\loop
	\StrChar{#1}{\value{i}}[\char]%
	\IsBoldNumber{\char}%
	\edef\result{\char}%
	\stepcounter{i}%
	\ifnum\value{i}<\numexpr\len+1
	\repeat
}
\newcommand{\ssss}[1]{[#1]}
\newcommand{\aaaa}[1]{\langle#1\rangle}
\newcommand{\ssaa}[1]{[#1\rangle}
\newcommand{\aass}[1]{\langle#1]}

\begin{document}

\hypersetup{pageanchor=false}
\begin{titlepage}

\begin{center}

\hfill UMN-TH-4316/24 \\
\hfill FTPI-MINN-24-07\\

\vskip 0.5in

{\huge \bfseries Momentum shift and on-shell constructible \vspace{4.5mm} \\ massive amplitudes}
\vskip .8in

{\large Yohei Ema,$^{1,2,a}$} \let\thefootnote\relax\footnote{$^a$ema00001@umn.edu}
{\large Ting Gao,$^{1,b}$}\footnote{$^b$gao00212@umn.edu}
{\large Wenqi Ke,$^{1,2,c}$}
\footnote{$^c$wke@umn.edu}
{\large Zhen Liu,$^{1,d}$}
\footnote{$^d$zliuphys@umn.edu}
\vspace{1.5mm}
\\
{\large Kun-Feng Lyu,$^{1,e}$}\footnote{$^e$lyu00145@umn.edu}
{\large Ishmam Mahbub$^{1,f}$}
\footnote{$^f$mahbu008@umn.edu}
\vskip .3in
\begin{tabular}{ll}
$^{1}$ & \!\!\!\!\!\emph{School of Physics and Astronomy, University of Minnesota, Minneapolis, MN 55455, USA}\\
$^{2}$ & \!\!\!\!\!\emph{William I. Fine Theoretical Physics Institute, School of Physics and Astronomy,}\\[-.15em]
& \!\!\!\!\!\emph{University of Minnesota, Minneapolis, MN 55455, USA}\\
\end{tabular}

\end{center}
\vskip .6in

\begin{abstract}
\noindent
We construct tree-level amplitude for massive particles using on-shell recursion relations
based on two classes of momentum shifts: an all-line transverse shift that deforms momentum by its transverse polarization vector, 
and a massive BCFW-type shift. 
We illustrate that these shifts allow us to correctly calculate four-point and five-point amplitudes in massive QED,
without an ambiguity associated with the contact terms that may arise from a simple ``gluing'' of lower-point on-shell amplitudes. 
We discuss various aspects and applicability of the two shifts, including the large-z behavior and complexity scaling. We show that there exists a ``good'' all-line transverse shift for all possible little group configurations of the external particles, which can be extended to a broader class of theories with massive particles such as massive QCD and theories with massive spin-1 particles. 
The massive BCFW-type shift enjoys more simplicity, 
but a ``good'' shift does not exist for all the spin states due to the specific choice of spin axis.

\end{abstract}

\end{titlepage}

\tableofcontents
\renewcommand{\thepage}{\arabic{page}}
\renewcommand{\thefootnote}{$\natural$\arabic{footnote}}
\setcounter{footnote}{0}
\hypersetup{pageanchor=true}

\section{Introduction}

On-shell recursion relation constructs tree-level scattering amplitudes focusing only on on-shell degrees of freedom.
Based on an introduction of shifts of external momenta, such as the Britto-Cachazo-Feng-Witten (BCFW) shift~\cite{Britto:2004ap,Britto:2005fq}, 
and the Cauchy theorem,
higher-point amplitudes are constructed recursively from on-shell lower-point amplitudes.
This method provides an alternative and, in certain cases, significantly more efficient way of evaluating scattering amplitudes.
It has enabled various proofs of fascinating properties of scattering amplitudes in various theories such as the Parke-Taylor formula~\cite{Parke:1986gb,Britto:2004ap,Britto:2005fq}, the color-kinematics duality 
and double-copy relations~\cite{Kawai:1985xq,Bern:2008qj,HenryTye:2010tcy,Bern:2010yg,Feng:2010my,Bjerrum-Bohr:2010pnr,Mafra:2011kj,Fu:2012uy} (see~\cite{Elvang:2013cua,Cheung:2017pzi,Travaglini:2022uwo} for reviews).

The on-shell method is best developed for amplitudes with only massless particles.
However, it is quite natural and tempting to extend this method to amplitudes with massive particles.
In this regard, Ref.~\cite{Arkani-Hamed:2017jhn} developed a systematic method of constructing 
three-point amplitudes for particles of any mass and spin
in a little group covariant way (see \textit{e.g.}~\cite{Kleiss:1985yh,Hagiwara:1985yu,Dittmaier:1998nn,Schwinn:2005pi,
Badger:2005zh,Badger:2005jv,Schwinn:2006ca,Ozeren:2006ft,Boels:2008du,Boels:2009bv,Boels:2010mj,Cohen:2010mi} 
for earlier works on the massive spinor-helicity formalism and recursion relation).
This method was used, \textit{e.g.}, to classify irreducible higher dimensional operators 
in effective field theories~\cite{Shadmi:2018xan,Durieux:2019eor,Ma:2019gtx,Durieux:2020gip,Balkin:2021dko,Bertuzzo:2023slg,Li:2022tec,Dong:2022mcv,Dong:2021yak,Li:2020gnx}, 
and to derive interactions of higher spin dark matter, black holes, massive supersymmetric amplitude, sterile neutrino and axion-like-particles~\cite{Chung:2018kqs,Arkani-Hamed:2019ymq,Herderschee:2019dmc, Herderschee:2019ofc,Falkowski:2020fsu,Chiodaroli:2021eug,Bertuzzo:2023slg,Song:2023jqm,Li:2021tsq,Li:2023wdz}.

On-shell construction of massive scattering amplitudes is often performed by ``gluing" lower-point amplitudes
without an explicit introduction of momentum shift.
While such ``gluing" procedures had been successful in understanding the pole structure of all possible amplitudes, the ambiguity of contact terms requires one to supply additional information like unitarity by hand in an unpredictable way, making it challenging to reproduce results in specific theories without referring to Feynman diagram results. 
For instance, the four-point $e^+ e^- \mu^+ \mu^-$ amplitude calculated in~\cite{Arkani-Hamed:2017jhn,Christensen:2022nja} 
by gluing the three-point amplitudes with the so-called $x$-factor (which we define later)
does not agree with the standard Feynman diagrammatic result, indicating the existence of a non-trivial contact term~\cite{Christensen:2022nja,Lai:2023upa}.
Ref.~\cite{Christensen:2022nja} attempts to construct the correct four-point amplitude, including the contact term, 
with different expressions of the three-point amplitude 
and obtained different results, finding the need for different contact terms for different expressions, which is not systematically understood.

In this paper, we introduce two momentum shifts, an all-line transverse shift and a massive BCFW-type shift, which allow us to properly construct massive scattering amplitudes,
including the contact terms,\footnote{
	In general, there are two types of contact terms: independent ones arising from \textit{e.g.}~higher dimensional operators that enter
	the calculation newly at a given multiplicity, and dependent ones related to lower-point amplitudes by \textit{e.g.}~gauge symmetry. 
	The former cannot be fully on-shell constructed  
	as the lower-point amplitudes miss certain physics inputs. SM, on the other hand, as a renormalizable gauge theory, is expected to be constructible (additional inputs on the Higgs potential might be needed). Hence, our main focus is on the latter types of contact terms. 
	See also the discussion around Eq.~\eqref{eq:constructibility}.
} without an ambiguity.
With these momentum shifts, we reproduce the correct four-point $e^+ e^- \mu^+ \mu^-$ amplitudes starting from the three-point amplitudes given in~\cite{Arkani-Hamed:2017jhn}.
Indeed, with the explicit momentum shift, 
intermediate photons that appear in our calculation always satisfy the on-shell kinematics.
Different expressions for the three-point amplitudes used in previous literature result in the same 
and correct amplitude, and therefore, no ambiguity related to the contact terms arises.
In this paper, we focus on four- and five-point amplitudes in massive QED as a first step, but we find that our momentum shift, in particular, the all-line transverse shift that we introduce below can be naturally extended to a broader class of theories, such as massive QCD and theories with massive spin-1 particles.
We will apply this method to the on-shell construction of the $WW$-scattering within the Standard Model (SM),
and reproduce the correct amplitude solely from the on-shell three-point amplitudes in a separate publication~\cite{Ema:2024rss}.

The rest of the paper is organized as follows. In Sec~\ref{sec:on-shell_general},
we give an overview of the on-shell recursion relation and our momentum shifts. 
We apply the general discussion there to the four- and five-point amplitudes in massive QED in Sec.~\ref{sec:QED}.
There, by introducing an explicit momentum shift, we reproduce the correct amplitude without an ambiguity
associated with the contact terms.
Finally, we summarize our findings and possible future directions in Sec.~\ref{sec:summary}.

\section{On-shell construction: general argument}
\label{sec:on-shell_general}

In this section, we provide an overview of the on-shell recursion relation and our proposed momentum shift.
The results in this section will be extensively used in our computation of the four- and five-point amplitudes
in massive QED in Sec.~\ref{sec:QED}.

\subsection{Recursion relation and factorization}

On-shell recursion relation constructs higher-point amplitude from lower-point on-shell information
(see~\cite{Elvang:2013cua,Cheung:2017pzi,Travaglini:2022uwo} for reviews).
The basis of on-shell recursion lies in shifting external momentum by arbitrary momentum with 
a complex parameter $z$ as
\begin{equation}
    \hat{p}_i (z)  = p_i + z  q_i.
 \label{eq2}
\end{equation}
We require that $\{\hat{p}_i(z)\}$ satisfy the total momentum conservation and 
keep the on-shell condition,\footnote{
	The latter may not be a strict requirement, 
	but to preserve the Ward identity at each pole, we demand it here.
}
\begin{align}
	\sum_i \hat{p}_i(z) = 0,
	\quad
	\hat{p}_i^2(z) = p_i^2 = m_i^2,
\end{align}
where we take all the momenta incoming by the crossing symmetry. 
These requirements put constraints on the possible choices of $q_i$.
After the shift, we think of an $n$-point amplitude as a complex function of $z$, $\hat{A}_n(z)$.
Focusing on tree-level amplitudes, $\hat{A}_n(z)$ has only poles in the complex $z$-plane
and hence is a meromorphic function,
and the Cauchy theorem tells us
\begin{align}
	 {A}_n &= \frac{1}{2\pi i}\oint_{z=0} \frac{dz}{z} \hat{A}_n(z)
  = - \sum_{\{z_I\}} \text{Res}\left[\dfrac{\hat{A}_n(z)}{z}\right] + B_\infty,
\end{align}
where we used that the original amplitude $A_n$ is given by $\hat{A}_n(z=0) = A_n$ 
in the first equality, and deformed the integration contour in the second equality,
picking up the poles $\{z_I\}$ and the boundary contribution $B_\infty$ at $\lvert z \lvert \rightarrow \infty$. Due to locality,
the pole corresponds to an intermediate particle going on-shell, $\hat{p}_I^2 (z_I) = m_I^2$,
and the residue factorizes into lower-point on-shell sub-amplitudes as 
\begin{align}
	A_n = -\sum_{z = z_I}\sum_\lambda\mathrm{Res}\left[\hat{A}^{(\lambda)}_{n-m+2}\frac{1}{z}\frac{1}{\hat{p}_I^2 - m_I^2}
	\hat{A}^{(-\lambda)}_{m}\right]
	+ B_\infty,
	\label{eq:recursion_general}
\end{align}
where $\lambda$ is the spin projection of the intermediate particle and $m \geq 3$ 
represents the number of external legs of the sub-amplitude. For a given factorization channel, 
if we divide external particles into two sets $L$ and $R$ representing left and right sub-amplitudes,
the internal momentum is given by
\begin{align}
    \hat{p}_I(z)  = p_I + z q_I = \ \sum_{i \in L} p_i + z  \sum_{i \in L} q_i.
\end{align}
Depending on $q_I^2 = 0$ or $q_I^2 \neq 0$, we have the poles at
\begin{align}
	\hat{p}_I^2 - m_I^2 = 0 ~~\Longrightarrow~~
	\begin{cases}
	z_I= -\dfrac{(p_I^2 - m_I^2)}{2p_I \cdot q_I}, & \mathrm{for}~q_I^2 = 0, \vspace{1.5mm} \\
	 z^{\pm}_I  =  \dfrac{1}{q_I^2}\left[
	-p_I \cdot q_I \pm \sqrt{(p_I \cdot q_I)^2 - (p_I^2 - m_I^2)q_I^2}\,\right],
	& \mathrm{for}~q_I^2 \neq 0.
	\end{cases}
\end{align}
In this paper, we will consider two distinct momentum shifts, an all-line transverse shift with $q_I^2 \neq 0$ 
and a massive BCFW-type shift with $q_I^2 = 0$.
The shift with $q_I^2 \neq 0$ is often dismissed in literature (see \textit{e.g.}~\cite{Cheung:2015cba})
due to the speculation that we have two solutions, with square roots, for each propagator
and this significantly complicates the computation.
However, as we see below, the poles always appear in pairs of $z_I^{\pm}$, which greatly simplifies the expression,
most notably canceling the square roots almost automatically, 
and hence, in most cases, it does not complicate the calculations significantly for tree-level amplitudes.

Eq.~\eqref{eq:recursion_general} suggests that we can construct higher-point amplitudes from lower-point on-shell amplitudes
if 
\begin{align}
	B_\infty = 0, ~~\mathrm{or}~~ \hat{A}(z) \to 0~~\mathrm{at}~~z \to \infty.
	\label{eq:constructibility}
\end{align}
This is indeed the condition for constructibility in~\cite{Benincasa:2007xk}.
Whether or not $B_\infty = 0$ depends both on the theory and the choice of the momentum shift $q_i$.
First,  to have $B_\infty = 0$, the theory should not have any \textit{independent} contact interaction
originating, \textit{e.g.}, from higher dimensional operators~\cite{Cohen:2010mi}.\footnote{
	A given interaction is said ``independent" if it is unrelated to lower-point interactions by any symmetry.
	This is not the case of, \textit{e.g.}, the self-interaction of non-abelian gauge bosons, as the gauge symmetry
	relates it to the lower-point amplitudes. 
} 
We can input this information, for instance, by restricting the mass dimension of couplings appearing in scattering amplitudes,
or by imposing the Ward identity at high energy.
In the absence of an independent contact interaction, there exists a ``good" momentum shift
that satisfies $B_\infty = 0$. 
The large-$z$ behavior of the amplitude 
depends on the choice of the momentum shift and a ``bad" momentum shift results in a finite boundary term $B_\infty \neq 0$,
even in the absence of an independent contact interaction.
Here, it may be useful to distinguish poles in $z$ and poles in the original momentum, $p_I^2 - m_I^2$.
Although the former arises from the latter of the original unshifted amplitude, they, in general, differ.
In particular, the dependent contact terms do not contribute to the residue of the latter (by definition),
but contribute to the former. This allows us to reproduce the dependent contact terms by the on-shell recursion relation method
with the ``good" momentum shift.
In the case of our interest, QED with massive fermions, we find two distinct good momentum shifts that both satisfy $B_\infty = 0$,
and introduce them in the following.

\subsection{Spinor formalism and momentum shift}
\label{subsec:shift}

In this subsection, we introduce two momentum shifts that equally work for our purpose.
Our momentum shift is based on the massive spinor formalism introduced in~\cite{Arkani-Hamed:2017jhn}.
In this formalism, we decompose the momentum $p$ of a massive particle in little group covariant form as
\begin{align}
	p_{a\dot{a}} = \lambda_{a}^I \tilde{\lambda}_{\dot{a}I},
	\quad
	I = 1,2,
\end{align}
where $I$ represents the little group SU(2) index, and dotted and undotted indices transform as (1/2,0) and (0,1/2) representation, respectively. 
We use bold spinors, $\lambda = \vert \textbf{p}\rangle$ and $\tilde{\lambda} = [\textbf{p}\vert$, 
to denote massive particles and keep the little group indices implicit unless ambiguity arises. 
These spinors satisfy the Dirac equation,
\begin{align}
    {p} \lvert \textbf{p} \rangle = m \lvert \textbf{p}]&, 
    \quad 
    {p} \lvert \textbf{p} ] = m \lvert \textbf{p}\rangle,
\end{align}
and therefore fermion wavefunctions can be constructed with $\vert \textbf{p}\rangle$ and $[\textbf{p}\vert$.
See App.~\ref{app:convention} for more details.

In the following, we discuss (i) the all-line transverse shift and (ii) the massive BCFW-type shift.
Even though the amplitude is little group covariant, we break the little group covariance
explicitly in our momentum shifts by performing different shifts for different combinations of external particles' spin states.
Remember that, in a massless theory, the BCFW shift should be performed
for a specific combination of helicity states since otherwise, the boundary term does not vanish~\cite{Britto:2005fq,Arkani-Hamed:2008bsc}.
We expect this also to be the case for massive amplitudes.
Indeed, both of our momentum shifts are based on specific choices of the spin axis/basis choice.\footnote{
	Ref.~\cite{Christensen:2022nja} instead attempted to keep the little group covariance manifest in their momentum shift,
	and failed to find a good momentum shift in QED.
}

\subsubsection*{All-line transverse shift}

Our first proposed proper massive momentum shift, the All-Line Transverse (ALT) shift, is presented in the helicity basis.
We expand the spinors as
\begin{align}
   \lvert \textbf{i} \rangle^I_a = \lvert i \rangle_a \delta^I_- + \lvert \eta_i \rangle_a \delta^{I}_+, \quad [ \textbf{i} \lvert^I_{\dot{a}} = [ i \lvert_{\dot{a}} \delta_+^I + [\eta_i \lvert _{\dot{a}} \delta^{I}_-,
\end{align}
where we use $I=\pm$ to indicate little group indices in the helicity basis. 
With this notation,  we have 
$\vert \mathbf{i}\rangle^{+} = \vert \eta_i \rangle,\vert \mathbf{i}\rangle^{-} = 
\vert i\rangle,[\mathbf{i}\vert^{+} = [i\vert,[\mathbf{i}\vert^{-} = [\eta_{i}\vert$,
and the momentum is given by
\begin{align}
	(p_i)_{a\dot{a}} = \vert i\rangle_a [i\vert_{\dot{a}} - \vert\eta_i\rangle_a [\eta_i\vert_{\dot{a}}.
\end{align}
The on-shell condition $p^2_i = m_i^2$ translates to
\begin{align}
	\langle i \eta_i \rangle = [i \eta_i] = m_i.
\end{align}
More details, including the explicit forms of the helicity basis as a two-component spinor, 
can be found in App.~\ref{app:convention}.

The ALT shift is defined according to the helicity of the external particle. For spin-$1/2$ and the transverse modes of spin-1, this shift is given by:
\begin{align}
	\begin{cases}
	\vert i \rangle \to \vert \hat{i}\rangle = \vert i\rangle + z c_i \vert \eta_i\rangle
	& \mathrm{for}~~I_i = +, \vspace{1mm} \\
	\vert i ] \to \vert \hat{i}] = \vert i] + z c_i \vert \eta_i]
	& \mathrm{for}~~I_i = -,
	\end{cases}\label{transverseshifts}
\end{align}
where $I_i$ is the helicity index of the particle $i$. For our purpose in this paper, we do not consider the longitudinal mode of a massive spin-1 particle, which will be treated in detail in a separate publication. Nevertheless, we note here that we can extend our momentum shift as
\begin{align}
	\begin{cases} 
	\vert i\rangle \to \vert \hat{i}\rangle = \vert i\rangle + z\dfrac{c_i}{2}\vert \eta_i \rangle, \vspace{1mm}\\
	[\eta_i\vert \to [\hat{\eta}_i\vert =  [\eta_i\vert - z\dfrac{c_i}{2}[i\vert,
	\end{cases}
	\mathrm{or}~~~
	\begin{cases}
	[i\vert \to [ \hat{i}\vert = [ i\vert + z\dfrac{c_i}{2}[ \eta_i \vert, \vspace{1mm}\\
	\vert\eta_i\rangle \to \vert\hat{\eta}_i\rangle = \vert\eta_i\rangle - z\dfrac{c_i}{2}\vert i\rangle,
	\end{cases}
	\mathrm{for}~~I_i = L,
	\label{eq:all_trans_long}
\end{align}
where ``$L$"  denotes the longitudinal mode of a massive spin-1 particle.
As one can see from Eqs.~\eqref{eq:fermion_wavefn} and~\eqref{eq:massive_pol_vectors_app}, 
these momentum shifts are defined such that they do not shift the fermion wavefunctions and polarization vectors of external particles up to spin-1,
including the longitudinal modes. Therefore,   wavefunctions and polarization vectors will be independent of $z$.
Now recall that the polarization vectors of the massive particles are given by
\begin{align}
	\epsilon^{(+)}_{a \dot{a}} = \sqrt{2}\frac{\vert \eta_i\rangle_a [ i \vert_{\dot{a}}}{m_i},
	\quad
	\epsilon^{(-)}_{a\dot{a}} = \sqrt{2}\frac{\vert i\rangle_a [ \eta_i \vert_{\dot{a}}}{m_i},
\end{align}
then we observe that the ALT shift modifies the momentum as
\begin{align}
	p_i \to \hat{p}_i = p_i + z \frac{c_i m_i}{\sqrt{2}} \epsilon_i^{(I_i)},
\end{align}
where we choose $I_i = \pm$.
In other words, we shift the momentum by the transverse polarization vector, as the name of the shift suggests. More precisely, for transverse modes, the polarization vector used in the momentum shift above matches with that of the external particle, whereas for longitudinal modes, we have an option of the helicity of the polarization vector in the shift. The first choice in Eq.~\eqref{eq:all_trans_long} corresponds to using the $+$~mode while the latter selects the $-$~mode in shifting longitudinal modes. Note that we cannot use the longitudinal polarization vector to shift the momentum as its norm does not vanish.\footnote{
	If one shifts the momentum by the longitudinal polarization vector, then $\hat{p}_i^2\neq p_i^2$ 
	due to the non-vanishing norm of $\epsilon^{(L)}_i$, so the on-shell condition no longer holds.
}

Now, we are going to show that the ALT shift indeed satisfies the on-shell condition and momentum conservation while leaving a free complex parameter identified as $z$. It is straightforward to see that the on-shell condition is intact, $\langle \hat{i} \eta_i \rangle = [ \hat{i} \eta_i ] = m_i$, because $\langle  \eta_i \eta_i \rangle = [ \eta_i\eta_i ] =0$.
Total momentum conservation imposes
\begin{align}
	0 = \sum_{i} c_i \epsilon_i^{(I_i)}.
	\label{eq:ci_momentum_cons}
\end{align}This seemingly puts four constraints, and hence naively has a non-trivial solution with $c_i \neq 0$
only when five or more legs are shifted. 
However, the transverse polarization vector does not have a temporal component,
which in the spinor-helicity language is expressed as
\begin{align}\label{eq:ci_eqs}
	\langle i \vert \gamma^0 \vert \eta_i] = \langle \eta_i \vert \gamma^0 \vert i] = 0.
\end{align}
This reduces the number of constraints to three and enables us to implement the ALT shift for four-point amplitude. 
For five- and higher-point amplitudes, we need not shift all the external lines to satisfy the momentum conservation.
However, the all-line shift makes the investigation of the large-$z$ behavior of the amplitude simpler.
Indeed, with the ALT shift, 
a simple dimensional analysis guarantees $B_\infty = 0$ in the case of massive QED, 
without even referring to Lagrangian, as we see below, 
and hence we focus on the all-line shift in the following.
We use this momentum shift to construct four- and five-point amplitudes recursively from three-point amplitudes in Sec.~\ref{sec:QED}.

One may worry that, in the ALT shift, it is cumbersome to derive the solution of $c_i$ that satisfies
the momentum conservation, as well as $z_I^\pm$ that satisfies the pole condition of intermediate particles.
However, we do not need explicit forms of $c_i$
in the actual computation, which allows us to avoid unnecessary complications. 
This is true, at least in the case of our interest, and we believe this is the case in general.
Eq.~\eqref{eq:ci_momentum_cons} is underdetermined and has an infinite number of solutions in addition
to the overall scaling (for $n\geq 5$),
yet the final result does not depend
on the specific choice of $c_i$, indicating that there exists a way of calculating the amplitudes without relying on
the explicit form of $c_i$.
Our computation of the four- and five-point amplitudes in QED in Sec~\ref{sec:QED} illustrates this point explicitly.

We also note that the ALT shift is equally defined for massless particles. Indeed, for massless particles, the momentum is expressed with a single pair of spinors, and we have the freedom of choosing arbitrary spinor $\vert \eta_i\rangle$ or $\vert \eta_i]$ to shift the momentum.\footnote{
	For massive particles, $\vert \eta_i\rangle$ and $\vert \eta_i]$ are defined by the momentum,
	but this is not the case for massless particles.
	Here we rather define $\vert \eta_i\rangle$ and $\vert \eta_i]$ through Eq.~\eqref{transverseshifts}
	for massless particles.
} For massless spin-1/2 particles, this choice is irrelevant, but
for massless spin-1 particles, we have two distinct options.
 If we pick them to be proportional to $\vert \xi_i \rangle$ and $\vert \xi_i]$, \textit{i.e.}~arbitrary spinors showing up in massless polarization vectors,
the shift is a natural massless limit of the ALT shift for massive spin-1 particles.
If we instead select them to be independent of $\vert \xi_i \rangle$ and $\vert \xi_i]$, the external polarization vectors
scale as $1/z$, indicating a better large-$z$ behavior, although satisfying the momentum conservation may require
additional care for four-point amplitudes. 
We take the former option in our discussion of large-$z$ behavior. However, the improved large-$z$ behavior of our shift in the presence of massless states may be useful for, \textit{e.g.}, constructing Compton amplitudes of non-renormalizable or higher spin theories.

We comment on the difference between our ALT shift and the massive all-line
shift introduced in~\cite{Cohen:2010mi}.
As shown in the reference, one has the freedom of selecting a massless reference spinor to define an on-shell massive spinor shift. After requiring the shift to keep external wavefunctions unchanged, we fix this reference spinor to be $\vert \eta_i\rangle$ or $\vert \eta_i]$ for each external leg in the helicity basis. Another method of achieving such all-line ``good'' shift, as done in the reference, is to work in a basis where all external massive spinors share a common light-like reference spinor. The common reference spinor guarantees that $q_i \cdot q_j = 0$ for any pair of momentum, which leads to the propagator scaling only as $z^{-1}$. This is often compensated by cancellations of $z$ in the numerators, which cannot be captured by the dimensional analysis alone and requires additional inputs such as ($q$-)helicity scaling. 
In the ALT shift, we have $q_i \cdot q_j \neq 0$ for $i\neq j$. The propagator scales as $z^{-2}$ consistent with the mass dimension, 
and a simple dimensional argument is enough to study the large-$z$ behavior, as we see below.
Lastly, the shift in~\cite{Cohen:2010mi} fails to reproduce amplitudes 
with, \emph{e.g.}, a certain combination of the external spin states in Yukawa theory, 
while the ALT shift works for the same amplitude.

\subsubsection*{Large-$z$ behavior}

Here, we will evaluate the large-$z$ behavior of ALT shift using dimensional analysis.
As argued in~\cite{Cheung:2015cba}, in general, $n$-point amplitudes can be decomposed as
\begin{align}
	A_n = \left(\sum_{\mathrm{diagrams}}g \times F\right) \times \prod_{\mathrm{vectors}} \epsilon \times \prod_{\mathrm{fermions}}u,\label{generalamp}
\end{align}
where $g$ collectively denotes the product of couplings, $F$ denotes Feynman diagrammatic contributions
with external polarization vectors and fermion wavefunctions stripped out,
$\epsilon$ schematically denotes the polarization vector of external particles, and $u$ represents the wavefunctions of external fermions which are of mass dimension $1/2$. 
We can further decompose $F$ into $F=N/D$, where  $D$ is  the product  of the denominator of all propagators, \textit{i.e.}~$D = \prod_{I}(\hat{p}_I^2 - m_I^2)$, with $I$ denoting the intermediate particle. The numerator $N$ contains the rest of the kinematic factors, including the momenta from derivative couplings, and factors of $\slashed{p}+m$ if the intermediate particle is a fermion.
An $n$-point amplitude has the mass dimension of $4-n$, and hence
\begin{align}
	4-n = [g] + [N] - [D] + \frac{1}{2}N_F,
\end{align}
where $N_F$ is the number of external fermions.

In the following, we focus on shifting all the external momenta.
This is not necessary to satisfy the total momentum conservation for $n \geq 5$,
but this greatly simplifies our study of the large-$z$ behavior.
Indeed, for the ALT shift and general kinematics of the external momenta,
the denominator of each propagator will scale as $z^2$.
This is not the case for non-all-line shifts, as discussed in~\cite{Cheung:2015cba}, since an intermediate propagator
can be a sum of momenta with only one of them shifted by a momentum $q_i$, with $q_i^2=0$ to satisfy the on-shell condition.
In such a case, $\hat{p}_I^2 = p_I^2 + 2z q_i \cdot p_I + z^2 q_i^2 = p_I^2 + 2z q_i \cdot p_I$ and hence the denominator scales as $z$
instead of $z^2$.
Furthermore, we will restrict our focus to spin-$1/2$ and spin-$1$ particles. 
Then, the ALT shift does not generate any $z$-dependence 
for external wavefunctions and polarization vectors, as we discussed above,
so that the large-$z$ behavior is entirely controlled by $F$. For the ALT shift,
the denominator of each  propagator scales as $z^2$, and hence we expect
\begin{align}
	D \sim z^{[D]}.
\end{align}
For the numerator, we have
\begin{align}
	N \sim z^{\gamma_N},
	\quad
	\gamma_N \leq [N],
\end{align}
because each shifted momentum contributes  with a factor of $z$  ($\hat{p}_i\sim z$). The upper bound is saturated when all mass dimension of $N$ is provided by $\hat{p}_i$.  
By combining these, we expect that at large $z$, the amplitude behaves as
\begin{align}
	\hat{A}_n \sim z^{\gamma},
	\quad
	\gamma = \gamma_N - [D] \leq [N] - [D] = 4-n - [g] - \frac{N_F}{2}.
	\label{eq:large_z}
\end{align}
Given that the coupling constants satisfy $[g] = 0$ in QED, it is  straightforward to prove that QED amplitudes have a good large-$z$ behavior under the ALT shift, namely   
$\hat{A}_n(z) \to 0$ at $\vert z\vert \to \infty$. Another remarkable feature of the ALT shift is that, for a renormalizable theory (\textit{i.e.}~$[g] \geq 0$) with spin$\leq1$, all the amplitudes with $n \geq 5$ 
are constructible by this shift. 
For $n = 4$, the amplitudes are constructible if they involve at least one fermion.

In a renormalizable theory and for amplitudes with spin-1 external states only, such as the $WW$-scattering, it is not obvious how to infer the large-$z$ behavior from Eq.~\eqref{eq:large_z}, because we would expect from this equation that $\gamma\leq0$
which allows $\gamma = 0$. However, as we shift the momenta by polarization vectors, the large-$z$ behavior is actually better than Eq.~\eqref{eq:large_z} thanks to the Ward identity, as long as we have at least one external transverse mode since the theory enjoys the gauge symmetry at high energy. Therefore, such theories are also constructible under the ALT shift. We will come back to this point in a separate publication~\cite{Ema:2024rss}.

\subsubsection*{Massive BCFW-type shift}

Next, we discuss another type of momentum shift, the massive BCFW-type (mBCFW) shift 
introduced in~\cite{Franken:2019wqr, Wu:2021nmq}. 
We note that any massive four-momentum can be decomposed as a sum of two null momenta,
\begin{align}
    p^{\mu}_i \ =  k^{\mu} +  \frac{m^2}{2 k \cdot r} r^{\mu} 
\end{align}
where $r^{\mu}$ is arbitrary reference vector. Under such decomposition, the spinors take the form
\begin{align}
  \label{eqeqeq} & \lvert \textbf{i} \rangle^1  =  \lvert k \rangle, \quad \lvert \textbf{i} \rangle^2  =  \frac{m}{\langle k r  \rangle} \lvert r \rangle,
     \quad \lvert \textbf{i} ]_1  =  \lvert k ], \quad \ \lvert \textbf{i} ]_2  =  \frac{m}{[ r k]} \lvert r ]
\end{align}
We then define 
$[\textbf{i}, \textbf{j} \rangle$ and $\langle \textbf{i}, \textbf{j} ]$ shift
by first writing momentum $p_i,p_j$ as a linear combination of null momentum $l_i,l_j$ 
which are co-planar to massive momentum
\begin{align}
    p_i  &=  l_i + \frac{m_i^2}{2 l_j \cdot l_i}  l_j, 
    \quad
    p_j  = l_j + \frac{m_j^2}{2 l_j\cdot l_i}l_i.
\end{align}
Then the shifted spinors are defined by
\begin{align}
	& [\textbf{i}, \textbf{j} \rangle \ \mathrm{shift}: \begin{cases}
	[ \hat{\textbf{i}}\lvert^2 =  [ \textbf{i} \lvert^2 - z[ \textbf{j} \lvert^2,   \vspace{1mm} \\
   \lvert \hat{\textbf{j}}\rangle^1 = \lvert \textbf{j} \rangle^1 + z\lvert \textbf{i} \rangle^1.
	\end{cases} 
\end{align}
The shift will preserve on-shell condition and conservation of total momentum. 
A ``good" shift,  which does not deform the external wavefunction, can be defined only for spin projection $(s_i,s_j) = (1,2)$ using  $[\textbf{i}, \textbf{j} \rangle$ shift and $(s_i,s_j) = (2,1)$ using $\langle \textbf{i}, \textbf{j} ]$ shift. Furthermore, these shifts will not produce any boundary term for four-point QED amplitude~\cite{Wu:2021nmq}. 
However, there does not exist a good mBCFW shift for $(s_i,s_j) = \{(1,1), (2,2)$\} for external fermion. 
This is in contrast to the massless case since massless amplitudes should have opposite helicity states. 
Otherwise, massless amplitudes will vanish due to the helicity selection rule. 
The mass breaks this rule, and we are allowed
to have non-zero amplitudes with, \emph{e.g.}, only one specific helicity.
To construct these amplitudes, we may invoke the little group covariance of the scattering amplitude.

\section{On-shell construction: QED}
\label{sec:QED}

Armed with the proper momentum shifts, we now construct four- and five-point amplitudes in QED using on-shell recursion relation.
In particular, we will illustrate the computation of $e^+ e^- \mu^+ \mu^-$ four-point amplitude 
with our momentum shifts, which resolves a question raised in~\cite{Christensen:2022nja}
regarding on-shell construction of massive amplitudes with internal photons.
The problem originates due to ``gluing" the amplitude without an explicit introduction of the momentum shift, as pole condition $\hat{s}=0$ is not properly taken into account.
After properly introducing the momentum shift, we show that no ambiguity related to the (dependent) contact terms 
arises, and we obtain the correct amplitude. We will then implement momentum shifts to obtain five-point $e^+ e^- \mu^+ \mu^- \gamma$ amplitude in order to present a general method of properly building higher-point massive amplitude.
\subsection{Four-point amplitude}

We now construct the four-point $e^+ e^- \mu^+ \mu^-$ amplitude by the on-shell recursion relation with our momentum shift.
Our starting point is Eq.~\eqref{eq:recursion_general}.
As we have argued, $B_\infty = 0$ for both the ALT and mBCFW shifts for this amplitude
(with a caveat that this holds only for a specific spin state for the latter),
and hence Eq.~\eqref{eq:recursion_general} reduces to
\begin{align}
	A_4 = -\sum_{z = z_I}\sum_\lambda\mathrm{Res}\left[\hat{A}^{(\lambda)}_{3}\frac{1}{z}\frac{1}{\hat{p}_I^2 - m_I^2}
	\hat{A}^{(-\lambda)}_{3}\right].
\end{align}
We call $e^+, e^-, \mu^+$ and $\mu^-$ as the particle $1,2,3$ and $4$, respectively.
Then, due to our flavor assignment, the pole exists only when $\hat{p}_I^2 = \hat{p}_{12}^2 = 0$, 
corresponding to the intermediate photon becoming on-shell,
where we use the short-hand notation $\hat{p}_{ij} = \hat{p}_i + \hat{p}_j$.
We note that, in general,
\begin{align}
	\frac{1}{\hat{p}_I^2 - m_I^2} = \begin{cases}
	 -\dfrac{1}{p_I^2 - m_I^2}\dfrac{z_I}{z-z_I} & \mathrm{for}~~q_I^2 = 0, \vspace{1mm} \\
	\dfrac{1}{p_I^2 - m_I^2}\dfrac{z_I^+ z_I^-}{(z-z_I^+)(z-z_I^-)} & \mathrm{for}~~q_I^2 \neq 0,
	\end{cases} \label{eq:prop}
\end{align}
and thus, we obtain
\begin{align}
	A_4(\mathbf{1234}) &= -\frac{1}{p_{12}^2}\frac{1}{z_{12}^+ - z_{12}^-}\sum_\lambda \left[
	z_{12}^-\left[\hat{A}_3(\mathbf{12}I^\lambda) \times \hat{A}_3(\mathbf{34}{-I}^{-\lambda})\right]_{z_{12}^+}
	- (z_{12}^+ \leftrightarrow z_{12}^-)
	\right],
	\label{eq:A4_helicity}
\end{align}
for the ALT shift, where $z_{ij}^\pm$ is the solution of $\hat{p}_{ij}^2 = 0$ and the subscript $z_{12}^\pm$ indicates
that the three-point amplitudes are evaluated at $z = z_{12}^\pm$, and
\begin{align}
	A_4(\mathbf{1234}) &= \frac{1}{p_{12}^2} 
	\sum_\lambda \left[\hat{A}_3(\mathbf{12}I^\lambda) \times \hat{A}_3(\mathbf{34}{-I}^{-\lambda})\right]_{z_{12}},
	\label{eq:A4_BCFW}
\end{align}
for the mBCFW shift, where $z_{ij}$ satisfies $\hat{p}_{ij}^2(z_{ij}) = 0$.
Note, in particular, that the latter expression is a simple ``gluing" of the three-point amplitudes, but with the crucial exception that
the momenta and spinors are shifted. 
In general, \textit{we cannot ignore these shifts}, or in other words, we are not allowed to ``unhat" (\textit{i.e.}~set $z$ to 0)
the amplitudes, as we see below.

To compute these expressions, we need the three-point amplitude.
Let us focus on $\hat{A}_3(\mathbf{12}I^\lambda)$ as the other one is constructed in the same way.
We need to evaluate the three-point functions only at the poles
and the intermediate photon is always on-shell in our computation.
As illustrated in~\cite{Arkani-Hamed:2017jhn}, to construct three point amplitude with two massive particles\footnote{with identical masses, which is true for QED,} and one massless particle, kinematics requires us to introduce arbitrary spinor $\xi$ since $\lvert \hat{I} \rangle$ and $\lvert \hat{I} ]$ are not independent. This is because $\hat{p}_1 \cdot \hat{p}_{I} = \hat{p}_2 \cdot \hat{p}_I=0$ 
and so $\langle \hat{I} \lvert \hat{\mathbf{1}} \lvert \hat{I} ] = \langle \hat{I} \lvert \hat{\mathbf{2}} \lvert \hat{I} ]=0$. 
We can build a massless spin-1 object called the $x$-factor as

\begin{align}
	\hat{x}_{12} = \frac{\langle \xi \vert (\hat{p}_1-\hat{p}_2) \vert \hat{I}]}{2m\langle \xi \hat{I}\rangle},
	\quad
	\hat{\tilde{x}}_{12} = \frac{[ \xi \vert (\hat{p}_1-\hat{p}_2) \vert \hat{I}\rangle}{2m[ \xi \hat{I}]},
\end{align}
where $m$ is the mass of the particles $1$ and $2$ and $\vert\xi \rangle$ is the reference spinor. Then the minimal coupling between massless spin-1 and two massive spin-1/2 particles is given by
\begin{equation}
\begin{split}
	&\hat{A}_3 (\textbf{1}\textbf{2}{I}^+)= \tilde{e}\hat{x}_{12}\langle \hat{\mathbf{1}} \hat{\mathbf{2}}\rangle
	= -\tilde{e}\frac{\langle \hat{\mathbf{1}} \xi \rangle [\hat{I}\hat{\mathbf{2}}]
	+ \langle \hat{\mathbf{2}}\xi \rangle[\hat{I}\hat{\mathbf{1}}]}{\langle \xi \hat{I}\rangle}, 
	\\
	&\hat{A}_3(\textbf{1}\textbf{2}I^-)= \tilde{e}\hat{\tilde{x}}_{12}
	[\hat{\mathbf{1}} \hat{\mathbf{2}}] =- \tilde{e}\frac{[ \hat{\mathbf{1}} \xi] \langle \hat{I} \hat{\mathbf{2}}\rangle
	+ [ \hat{\mathbf{2}} \xi]\langle \hat{I} \hat{\mathbf{1}}\rangle}{[ \xi \hat{I}]},
	\label{3pt}
 \end{split}
\end{equation}
where $\tilde{e}= \sqrt{2}e$ denotes the coupling constant, and
the last expressions can be identified as the expected minimal coupling, $e\bar{u} \slashed{\epsilon}^{\pm}v$. Furthermore, we used $\hat{p}_I^2 = 0$ as well as the Schouten identity in the second equality and emphasize that the equality holds only for on-shell photons. In the following discussion, we use the second expression in the factorization channel, and use momentum shift to construct the amplitude. One can refer to the appendix for further details regarding on-shell construction starting from x-factor three-point amplitude. Now,  
we then obtain by using the second expression of Eq.~\eqref{3pt}
\begin{align}
	\sum_{\lambda}\hat{A}_3(\mathbf{12}I^\lambda) \times \hat{A}_3(\mathbf{34}{-I}^{-\lambda})
	&=\tilde{e}^2\left(\frac{\langle \xi \hat{\textbf{2}} \rangle [ \hat{\textbf{1}} \hat{I}]}{ \langle \xi \hat{I}\rangle} + \frac{\langle  \hat{\textbf{1}} \xi \rangle [\hat{I} \hat{\textbf{2}} ]}{\langle \xi \hat{I}\rangle} \right) \left( \frac{[ \xi {\hat{\textbf{4}}} ] \langle \hat{\textbf{3}} -\hat{I} \rangle}{ [ \xi -\hat{I} ] } + \frac{[  \hat{\textbf{3}} \xi ] \langle -\hat{I} {\hat{\textbf{4}}} \rangle}{[ \xi -\hat{I} ] } \right ) + \left([...] \leftrightarrow \langle ... \rangle\right), \label{eq:four-point-pole}
\end{align}
where the second term indicates exchanging the angle and square brackets of the first term,
and it is understood that we evaluate the above product at the pole, \textit{i.e.}~$z= z_{12}^\pm$ or $z_{12}$.
Using Schouten identity and the on-shell condition of the external legs, we can rewrite the above as
(see App.~\ref{app:details_4pt} for details)
\begin{align}
	\sum_{\lambda}\hat{A}_3(\mathbf{12}I^\lambda) \times \hat{A}_3(\mathbf{34}{-I}^{-\lambda})
	= \tilde{e}^2\left(
	\langle \hat{\textbf{1}} \hat{\textbf{3}} \rangle [\hat{\textbf{2}}\hat{\textbf{4}}] + \langle \hat{\textbf{1}}\hat{\textbf{4} }\rangle [\hat{\textbf{2}}\hat{\textbf{3}}] +\langle \hat{\textbf{2}}\hat{\textbf{3}} \rangle [\hat{\textbf{1}}\hat{\textbf{4}}] +\langle \hat{\textbf{2}} \hat{\textbf{4} }\rangle [\hat{\textbf{1}}\hat{\textbf{3}}]  \right).
	\label{eq:A3prod_hat}
\end{align}
As we have explained in Sec.~\ref{subsec:shift}, both ALT and mBCFW shifts are chosen so that  
we do not shift the wavefunctions of the external fermions.
In other words, our momentum shift is such that the spinors with un-contracted little group indices appearing
in the scattering amplitudes, which arise from the external wavefunctions and/or polarization vectors, are not shifted.
Therefore, the above expression is equivalent to
\begin{align}
	\sum_{\lambda}\hat{A}_3(\mathbf{12}I^\lambda) \times \hat{A}_3(\mathbf{34}{-I}^{-\lambda})
	= \tilde{e}^2\left[
	\langle {\textbf{1}} {\textbf{3}} \rangle [{\textbf{2}}{\textbf{4}}] + \langle {\textbf{1}}{\textbf{4} }\rangle
	[{\textbf{2}}{\textbf{3}}] +\langle {\textbf{2}}{\textbf{3}} \rangle 
	[{\textbf{1}}{\textbf{4}}] +\langle {\textbf{2}} {\textbf{4} }\rangle [{\textbf{1}}{\textbf{3}}]  \right],
	\label{eq:A3prod_unhat}
\end{align}
where we now do not have hats on the spinors.
Notice that, although somewhat implicit in the above computation,
our momentum shift actually differs for different combinations of the external helicity/spin states,
and is possible only for a specific spin state for the mBCFW shift.
By inserting this expression to Eq.~\eqref{eq:A4_helicity} or Eq.~\eqref{eq:A4_BCFW}, we obtain
\begin{align}
	A_4 = \frac{\tilde{e}^2}{p_{12}^2}
	\left[
	\langle {\textbf{1}} {\textbf{3}} \rangle [{\textbf{2}}{\textbf{4}}] + \langle {\textbf{1}}{\textbf{4} }\rangle
	[{\textbf{2}}{\textbf{3}}] +\langle {\textbf{2}}{\textbf{3}} \rangle 
	[{\textbf{1}}{\textbf{4}}] +\langle {\textbf{2}} {\textbf{4} }\rangle [{\textbf{1}}{\textbf{3}}] 
	\right],
	\label{eq:A4_correct}
\end{align}
for both ALT and mBCFW shifts.
This agrees with the amplitude obtained in the standard Feynman diagrammatic method.
We thus reproduce the correct four-point $e^+ e^- \mu^+ \mu^-$ amplitude, without the contact term ambiguity, 
by the on-shell recursion relation.

Now that we have reproduced the correct amplitude, we take a closer look at the computation of four-point $e^+ e^- \mu^+ \mu^-$ amplitude using factorization in the existing literature.
In~\cite{Arkani-Hamed:2017jhn,Christensen:2022nja}, this amplitude was calculated in the massive spinor-helicity formalism
by ``gluing" the three-point amplitudes, expressed by the $x$-factor, 
\emph{without} introducing the momentum shift. 
As we have noted, the second equality in Eq.~\eqref{3pt} requires $\hat{p}_I^2 = 0$ 
and this relation does not hold in general without the momentum shift.
This introduces an ambiguity in evaluating the three-point amplitudes with different expressions,
corresponding to an ambiguity in contact terms.
In particular, the amplitude was calculated in~\cite{Arkani-Hamed:2017jhn} based on the $x$-factor as
\begin{align}
    \left.A_4\right\vert_{\mathrm{AHH}} = 
    \tilde{e}^2 \frac{\langle \textbf{1} \textbf{2} \rangle \langle \textbf{3} \textbf{4} \rangle}{m s} (p_1 - p_2)\cdot p_3,
\end{align}
where $s = p_{12}^2$ and the electron and muon masses are equally taken as $m$.
The authors in~\cite{Christensen:2022nja} gave another expression,
\begin{align}
     \left.A_4\right\vert_{\mathrm{CDFHM}} = 
     \dfrac{\tilde{e}^2}{2 m_e m_\mu s}  \bigg[ \brr{u - t + 2 m_e^2 + 2 m_\mu^2} \squf{\textbf{1} \textbf{2}} \squf{\textbf{3} \textbf{4}} 
    + 2 \brr{ \squf{\textbf{1} \textbf{2}} [\textbf{3} |\textbf{2}\textbf{1}| \textbf{4}] 
    +  \squf{\textbf{3} \textbf{4}}[\textbf{1}|\textbf{4}\textbf{3}\lvert \textbf{2}]  } \bigg].
	\label{christensen}
\end{align}
Although these expressions correctly reproduce the residue at $s = 0$, 
they differ from the expected result~\eqref{eq:A4_correct} for general kinematics, away from $s = 0$.
We see that introducing an explicit momentum shift solves this discrepancy and gives   automatically the correct answer.
Indeed, by using the first expression in Eq.~\eqref{3pt}, we obtain with our momentum shift
\begin{align}
	\sum_{\lambda}&\hat{A}_3(\mathbf{12}I^\lambda) \times \hat{A}_3(\mathbf{34}{-I}^{-\lambda})
	\nonumber \\
	&= \frac{\tilde{e}^2}{2 m_e m_\mu}  \left[ \brr{\hat{u} - \hat{t} + 2 m_e^2 + 2 m_\mu^2} \squf{\hat{\textbf{1}} \hat{\textbf{2}}} \squf{\hat{\textbf{3}} \hat{\textbf{4}}} 
    + 2 \brr{ \squf{\hat{\textbf{1}} \hat{\textbf{2}}} \hat{[\textbf{3}} |\hat{\textbf{2}} \hat{\textbf{1}}| \hat{\textbf{4}}] +  \squf{\hat{\textbf{3}} \hat{\textbf{4}}}[\hat{\textbf{1}}|\hat{\textbf{4}} \hat{\textbf{3}}\lvert \hat{\textbf{2}}]  } \right],
    \label{christensen2}
\end{align}
where the product is to be evaluated at the pole $z = z_{12}^\pm$ or $z_{12}$.
This is analogous to Eq.~\eqref{christensen}, but with a crucial difference that the spinors and momenta are shifted (or hatted), and we cannot simply unhat this expression using ALT shift.
We can rewrite it as
\begin{align}
	\sum_{\lambda}&\hat{A}_3(\mathbf{12}I^\lambda) \times \hat{A}_3(\mathbf{34}{-I}^{-\lambda})
	\nonumber \\
	&=\tilde{e}^2 
	 \bigg[ \frac{\hat{s}}{2m_em_{\mu}} \brr{ \squf{\hBf{1} \hBf{2}} \squf{\hBf{3} \hBf{4}} - 2 \squf{\hBf{1} \hBf{4}} \squf{\hBf{2} \hBf{3}}} +
   \squf{\hBf{1} \hBf{3}} \angf{\hBf{2} \hBf{4}}  + \squf{\hBf{2} \hBf{4}} \angf{\hBf{1} \hBf{3}} + \squf{\hBf{1} \hBf{4}} \angf{\hBf{2} \hBf{3}}  + \squf{\hBf{2} \hBf{3}} \angf{\hBf{1} \hBf{4}} \bigg].
   \label{christensen3}
\end{align}
Note that the relation $\hat{p}_{12}^2 =0$ has not been used to derive \eqref{christensen3} from \eqref{christensen2}. 
Now, by applying the pole condition $\hat{s}=\hat{p}_{12}^2=0$, it reduces to Eq.~\eqref{eq:A3prod_hat} and hence
to Eq.~\eqref{eq:A3prod_unhat}.\footnote{
	A similar expression was shown in~\cite{Lai:2023upa}, but again without performing the momentum shift.
	The authors pointed out that once one ignores the terms proportional to $p_{12}^2$, it reproduces the correct amplitude.
	In our computation with the momentum shift, this assumption is proven automatically 
	since we need to evaluate the three-point amplitudes only at the pole $\hat{p}_{12}^2 = 0$.
We solve the puzzle systematically within the massive helicity amplitude framework. 
} In other words, the amplitude with the explicit introduction of the momentum shift always yields the right answer,
regardless of the different expressions  that one starts with.

\subsection{Five-point amplitude}
\label{sec:five_point_amp}

\begin{figure}[t]
    \centering
    \includegraphics[width=0.85\textwidth]{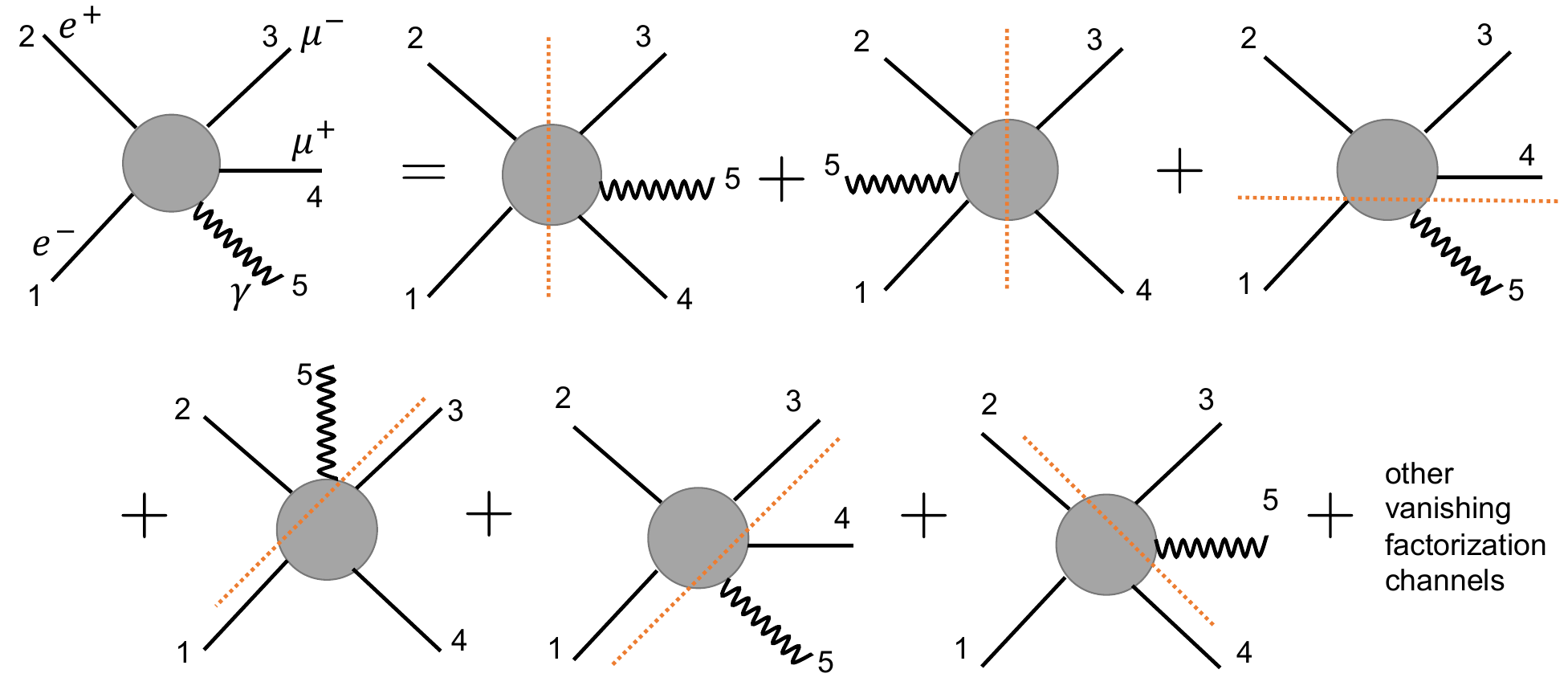}
    \caption{Factorization channels for $e^+e^-\mu^+\mu^-\gamma$ amplitude.   }
    \label{fig:diagram}
\end{figure}

We now extend our analysis to construct five-point amplitude, $e^+e^-\mu^+\mu^-\gamma$, 
using both ALT shift and BCFW-type shift.
For the ALT shift, we need to sum over six poles, which diagrammatically take the form of Fig.~\ref{fig:diagram}.
We will label $e^-,e^+,\mu^-,\mu^+,\gamma$ respectively as 1, 2, 3, 4, 5, and take the photon helicity to be negative for definiteness
(the positive helicity case is completely analogous). 
Let us first consider the factorization channel $\hat{p}_{35}^2 = m_\mu^2$ (6th diagram) 
for which we can write the product of the lower-point amplitudes as
 \begin{align}
\sum_{\lambda}\hat{A}_4(\mathbf{124}I^\lambda) \times \hat{A}_3(\mathbf{3}5^-{-I}^{-\lambda})\Big|_{\hat{p}_{35}^2 = m_\mu^2} =&\frac{\tilde{e}^3}{\hat{p}_{12}^2}\left(
	\langle {\hat{\textbf{1}}} \hat{{\textbf{I}}}^I \rangle [\hat{{\textbf{2}}}\hat{{\textbf{4}}}] + \langle \hat{{\textbf{1}}}\hat{{\textbf{4}}}\rangle
	[\hat{{\textbf{2}}}\hat{{\textbf{I}}}^I] +\langle \hat{{\textbf{2}}}\hat{{\textbf{I}}}^I \rangle 
	[\hat{{\textbf{1}}}\hat{{\textbf{4}}}] +\langle \hat{{\textbf{2}}} \hat{{\textbf{4} }}\rangle [\hat{{\textbf{1}}}\hat{{\textbf{I}}}^I]  \right) \tilde{\hat{x}}_{I3}[-\hat{\textbf{I}}_I\hat{\textbf{3}}] \nonumber \\
 =& \frac{\tilde{e}^3}{\hat{p}_{12}^2} \bigg[ \frac{\langle \hat{5} \lvert \hat{p}_4 \hat{p}_3 \lvert \hat{5} \rangle }{2\hat{p}_4 \cdot \hat{p}_5} \left( 	\langle {\hat{\textbf{1}}} \hat{{\textbf{3}}} \rangle [\hat{{\textbf{2}}}\hat{{\textbf{4}}}] + \langle \hat{{\textbf{1}}}\hat{{\textbf{4}}}\rangle
	[\hat{{\textbf{2}}}\hat{{\textbf{3}}}] +\langle \hat{{\textbf{2}}}\hat{{\textbf{3}}} \rangle 
	[\hat{{\textbf{1}}}\hat{{\textbf{4}}}] +\langle \hat{{\textbf{2}}} \hat{{\textbf{4} }}\rangle [\hat{{\textbf{1}}}\hat{{\textbf{3}}}] \right) \nonumber \\
& \qquad  + [\hat{\textbf{1}} \hat{\textbf{4}}] \langle \hat{\textbf{2}} \hat{ 5} \rangle \langle \hat{\textbf{3}} \hat{5} \rangle + [\hat{\textbf{2}}\hat{\textbf{4}}] \langle \hat{\textbf{1}} \hat{5} \rangle \langle \hat{\textbf{3}} \hat{5} \rangle\bigg],
\label{eq:p3p5}
 \end{align}
 where $\hat{p}_I = - \hat{p}_{124}= \hat{p}_{35}$. Similarly, for the factorization channel $\hat{p}_{45}^2 = m_\mu^2 $ (5th diagram), we find the amplitude product
 \begin{align}
\sum_{\lambda}\hat{A}_4(\mathbf{123}I^\lambda) \times \hat{A}_3(\mathbf{4}5^-{-I}^{-\lambda}) \Big|_{\hat{p}_{45}^2 = m_\mu^2}=& \frac{\tilde{e}^3}{\hat{p}_{12}^2} \bigg[ \frac{\langle \hat{5} \lvert \hat{p}_4 \hat{p}_3 \lvert \hat{5} \rangle }{2\hat{p}_3 \cdot \hat{p}_5} \left( 	\langle {\hat{\textbf{1}}} \hat{{\textbf{3}}} \rangle [\hat{{\textbf{2}}}\hat{{\textbf{4}}}] + \langle \hat{{\textbf{1}}}\hat{{\textbf{4}}}\rangle
	[\hat{{\textbf{2}}}\hat{{\textbf{3}}}] +\langle \hat{{\textbf{2}}}\hat{{\textbf{3}}} \rangle 
	[\hat{{\textbf{1}}}\hat{{\textbf{4}}}] +\langle \hat{{\textbf{2}}} \hat{{\textbf{4} }}\rangle [\hat{{\textbf{1}}}\hat{{\textbf{3}}}] \right) \nonumber \\
& \qquad  - [\hat{\textbf{1}} \hat{\textbf{3}}] \langle \hat{\textbf{2}} \hat{ 5} \rangle \langle \hat{\textbf{4}} \hat{5} \rangle - [\hat{\textbf{2}}\hat{\textbf{3}}] \langle \hat{\textbf{1}} \hat{5} \rangle \langle \hat{\textbf{4}} \hat{5} \rangle\bigg].
\label{eq:p4p5}
 \end{align}
For the factorization channel $\hat{p}_{12}^2=0$ (1st diagram) where we have internal photon, the amplitude product is given by 
\begin{align}
&\sum_{\lambda}\hat{A}_3(\mathbf{12}I^\lambda) \times \hat{A}_4(\mathbf{34}5^-{-I}^{-\lambda}) \Big|_{\hat{p}_{12}^2=0}=\tilde{e}^3 \bigg[ \frac{\langle \hat{5} \lvert \hat{p}_4 \hat{p}_3 \lvert \hat{5} \rangle}{4  \hat{p}_3 \cdot \hat{p}_5 \hat{p}_4 \cdot \hat{p}_5} \left( \langle \hat{\textbf{1}}  \hat{\textbf{4}} \rangle [\hat{\textbf{2}} \hat{\textbf{3}} ]+\langle \hat{\textbf{2}}  \hat{{\textbf{4}}} \rangle [\hat{\textbf{1}} \hat{\textbf{3}} ]+\langle \hat{\textbf{1}} \hat{\textbf{3}} \rangle [\hat{\textbf{2}} \hat{\textbf{4}}]+ \langle \hat{\textbf{2}} \hat{\textbf{3}} \rangle [ \hat{\textbf{1}} \hat{\textbf{4}} ] \right) \nonumber \\\
     & \qquad  + \frac{1}{2 \hat{p}_3 \cdot \hat{p}_5} ( [\hat{\textbf{1}} \hat{\textbf{4}}]\langle  \hat{\textbf{2}} \hat{5}\rangle \langle \hat{\textbf{3}} \hat{5}\rangle+ [\hat{\textbf{2}} \hat{\textbf{4}}]\langle  \hat{\textbf{1}} \hat{5}\rangle \langle \hat{\textbf{3}} \hat{5}\rangle ) 
        - \frac{1}{2 \hat{p}_4\cdot\hat{p}_5} ( [\hat{\textbf{1}} \hat{\textbf{3}}]\langle  \hat{\textbf{2}} \hat{5}\rangle \langle \hat{\textbf{4}} \hat{5}\rangle+ [\hat{\textbf{2}} \hat{\textbf{3}} ]\langle  \hat{\textbf{1}}  \hat{5}\rangle \langle \hat{\textbf{4}} \hat{5} \rangle ) \bigg].
        \label{eq:s_12}
\end{align}
It should be noted that there are multiple equivalent expressions for these amplitude products that differ by the pole condition. The above expression is chosen since its functional form matches that of other poles and when we sum over all factorization channels, the $z$-dependence cancels off. We will now explicitly show this cancellation of $z$-dependence after summing over factorization channels. From Eq.~\eqref{eq:prop}, we note that the propagators for this case can be written as
\begin{align}
    \frac{1}{2\hat{p}_3  \cdot \hat{p}_5} &= \frac{1}{2 p_3 \cdot p_5} \frac{z_{35}^+ }{(z-z_{35}^+)} \frac{z_{35}^-}{(z-z_{35}^-)},\\
    \frac{1}{2\hat{p}_4  \cdot \hat{p}_5} &= \frac{1}{2 p_4 \cdot p_5} \frac{z_{45}^+ }{(z-z_{45}^+)}\frac{z_{45}^-}{(z-z_{45}^-)}, \\
    \frac{1}{\hat{p}_{12}^2} &= \frac{1}{p_{12}^2} \frac{z_{12}^+ }{(z-z_{12}^+)} \frac{z_{12}^-}{(z-z_{12}^-)}.
\end{align}
Now consider terms in different factorization channels that share the same set of propagators $\{ \hat{P}\}$ and the same functional form $N(z)$. Then, the sum over different channels can be expressed as
\begin{align}
   \frac{1}{P} \sum_{k=1}^I N(z_k) \prod_{i=1,i\neq k}^I \frac{z_i}{z_k - z_i}=   \frac{N}{P} \sum_{k=1}^I \sum_{j=0}^J a_j z^j_k \prod_{i=1,i\neq k}^I \frac{z_i}{z_k - z_i} =   \frac{N}{P}  \sum_{j=0}^J a_j f_j,
\end{align}
where 
\begin{align}
    f_j = \sum_{k=1}^I  z^j_k \prod_{i=1,i\neq k}^I \frac{z_i}{z_k - z_i}.
\end{align}
Here $P$ denotes the product of all unhatted propagators in the set, and $\{ z_i \}$ is the set of all possible poles for the propagators. In the second equality, we expanded $N(z)$ in powers of $z$ and factored out the constant term $N$ (\textit{i.e.}~set $a_0=1$). Now one can show that $f_0 = (-1)^{I+1}$, and for $1 \leq j < I$, $f_j=0$. 
Thus, we observe that if different factorization channels share the same functional form with a common set of propagators, by summing over all the resides belonging to this set of propagators, we can drop off the $z$ dependence if $J<I$. Finally, the contributions from other poles can be found using symmetry $(1,2) \leftrightarrow (3,4)$, and we find the five-point amplitude, 
\begin{align}
   & A_5 (\textbf{1},\textbf{2},\textbf{3},\textbf{4},5^-) = \frac{\tilde{e}^3}{p_{12}^2} \bigg[ \frac{\langle {5} \lvert {p}_4 {p}_3 \lvert {5} \rangle}{4  {p}_3 \cdot {p}_5 {p}_4 \cdot {p}_5} \left( \langle {\textbf{1}}  {\textbf{4}} \rangle [{\textbf{2}} {\textbf{3}} ]+\langle {\textbf{2}}  {{\textbf{4}}} \rangle [{\textbf{1}} {\textbf{3}} ]+\langle {\textbf{1}} {\textbf{3}} \rangle [{\textbf{2}} {\textbf{4}}]+ \langle {\textbf{2}} {\textbf{3}} \rangle [ {\textbf{1}} {\textbf{4}} ] \right) \nonumber \\\
     & \qquad  + \frac{1}{2 {p}_3 \cdot {p}_5} ( [{\textbf{1}} {\textbf{4}}]\langle  {\textbf{2}} {5}\rangle \langle {\textbf{3}} {5}\rangle+ [{\textbf{2}} {\textbf{4}}]\langle  {\textbf{1}} {5}\rangle \langle {\textbf{3}} {5}\rangle ) 
        - \frac{1}{2 {p}_4\cdot{p}_5} ( [{\textbf{1}} {\textbf{3}}]\langle  {\textbf{2}} {5}\rangle \langle {\textbf{4}} {5}\rangle+ [{\textbf{2}} {\textbf{3}} ]\langle  {\textbf{1}}  {5}\rangle \langle {\textbf{4}} {5} \rangle ) \bigg] + (1,2) \leftrightarrow (3,4),
        \label{eq:five-point}
\end{align}
where all the $z$-dependence cancels out as discussed.
Further details of the calculation can be found in App.~\ref{app:details_5pt}. We can derive identical results using massive-massless BCFW shift \cite{Ballav:2020ese} 
\begin{align}
	&[ \hat{5} \lvert  =    [ {5} \lvert + z\langle 5 \lvert  p_4,
        \quad
    \lvert \hat{\textbf{4}} \rangle  =    \lvert {\textbf{4}} \rangle - z\langle 5 \textbf{4} \rangle  \lvert 5 \rangle.
\end{align}
Using this shift, we will hit four poles (diagram 2, 3, 4, 6). Consider $\hat{p}_{35}^2 = m_\mu^2$ channel for which we find $z_{35} = - \frac{2p_3 \cdot p_5}{\langle 5 \lvert p_4 p_3 \lvert 5 \rangle} $. Then residue using $[ 5 \textbf{4} \rangle$ shift is given by
\begin{align}
  &\sum_{\lambda}\hat{A}_4(\mathbf{124}I^\lambda) \times \hat{A}_3(\mathbf{3}5^-{-I}^{-\lambda})\Big|_{\hat{p}_{35}^2 = m_\mu^2} \nonumber \\ & \quad = \frac{\tilde{e}^3}{{p}_{12}^2} \bigg[ \frac{\langle {5} \lvert  \hat{p}_4 {p}_3\lvert {5} \rangle }{2{p}_4 \cdot {p}_5} \left( 	\langle {{\textbf{1}}} {{\textbf{3}}} \rangle [{{\textbf{2}}}{{\textbf{4}}}] + \langle {{\textbf{1}}}\hat{{\textbf{4}}}\rangle
	[{{\textbf{2}}}{{\textbf{3}}}] +\langle {{\textbf{2}}}{{\textbf{3}}} \rangle 
	[{{\textbf{1}}}{{\textbf{4}}}] +\langle {{\textbf{2}}} \hat{{\textbf{4} }}\rangle [{{\textbf{1}}}{{\textbf{3}}}] \right) + [{\textbf{1}} {\textbf{4}}] \langle {\textbf{2}} { 5} \rangle \langle {\textbf{3}} {5} \rangle + [{\textbf{2}}{\textbf{4}}] \langle {\textbf{1}} {5} \rangle \langle {\textbf{3}} {5} \rangle\bigg] \nonumber \\
& \quad =  \frac{\tilde{e}^3}{{p}_{12}^2} \bigg[ \frac{\langle {5} \lvert  {p}_4 {p}_3 \lvert {5} \rangle }{2{p}_4 \cdot {p}_5} \left( 	\langle {{\textbf{1}}} {{\textbf{3}}} \rangle [{{\textbf{2}}}{{\textbf{4}}}] + \langle {{\textbf{1}}}{{\textbf{4}}}\rangle
	[{{\textbf{2}}}{{\textbf{3}}}] +\langle {{\textbf{2}}}{{\textbf{3}}} \rangle 
	[{{\textbf{1}}}{{\textbf{4}}}] +\langle {{\textbf{2}}} {{\textbf{4} }}\rangle [{{\textbf{1}}}{{\textbf{3}}}] \right) \nonumber \\
 & \qquad \qquad  + [{\textbf{1}} {\textbf{4}}] \langle {\textbf{2}} { 5} \rangle \langle {\textbf{3}} {5} \rangle + [{\textbf{2}}{\textbf{4}}] \langle {\textbf{1}} {5} \rangle \langle {\textbf{3}} {5} \rangle - \frac{p_3 \cdot p_5}{p_4 \cdot p_5} \left (  [{\textbf{1}} {\textbf{3}}] \langle {\textbf{2}} { 5} \rangle \langle {\textbf{4}} {5} \rangle +[{\textbf{2}}{\textbf{3}}] \langle {\textbf{1}} {5} \rangle \langle {\textbf{4}} {5} \rangle  \right) \bigg].
\end{align}
We notice that the residue of this channel contains other poles, as one might expect. The sum over the rest of the poles gives the above expression with the exchange $(1,2) \leftrightarrow (3,4)$, which we can argue from symmetry and thus derive the result in Eq.~\eqref{eq:five-point}. To explicitly calculate the residue of three other poles, one would write down Eq.~(\ref{eq:p3p5}$-$\ref{eq:s_12}) with $(1,2) \leftrightarrow (3,4)$. Then, using arguments similar to the ALT shift, one can drop the $z$-dependent factors since they will cancel off after summing over all channels.

We may note that the amplitudes correctly satisfy the leading and sub-leading soft theorem~\cite{Low:1954kd,Low:1958sn,Weinberg:1965nx,Cachazo:2014fwa,Casali:2014xpa,He:2014cra,Lysov:2014csa,Falkowski:2020aso,McLoughlin:2022ljp}.
Indeed, the leading and sub-leading soft-factors of the amplitude are given by
\begin{align}
S_-^{(0)} &=  \sum_{i=1}^{n} Q_i \frac{\epsilon_{-\mu} (q)p_i^{\mu}}{q \cdot p_i} = \tilde{e} \frac{\langle 5 \lvert p_4 p_3 \lvert 5 \rangle}{4 (p_3 \cdot p_5) (p_4 \cdot p_5)}+ (1,2) \leftrightarrow (3,4),
	\\
    S_-^{(1)} &=  -i  \sum_{i=1}^{n} Q_i \frac{\epsilon_{-\mu} (q)q_{\nu} J_i^{\mu \nu}}{q \cdot p_i} =\tilde{e} \left(\frac{\langle 5 \textbf{3}^L \rangle \langle 5  \lvert^{\alpha}\frac{\partial}{\partial\lambda^{\alpha L}_3} }{2p_3 \cdot p_5} -  \frac{\langle 5 \textbf{4}^L \rangle \langle 5  \lvert^{\alpha}\frac{\partial}{\partial\lambda^{\alpha L}_4} }{2p_4 \cdot p_5} \right) + (1,2) \leftrightarrow (3,4),
\end{align}
and the five-point amplitude takes the form
\begin{align}
    {A}_5 (\textbf{1},\textbf{2},\textbf{3},\textbf{4},5^-) =  \left [ \sum_{i=0}^{1 }  S_{-}^{(i)} \right] A_4(\textbf{1},\textbf{2},\textbf{3},\textbf{4}).
\end{align}

\section{Summary and outlook}
\label{sec:summary}
Constructing higher point amplitudes from lower point building blocks is an essential feature of on-shell methods. With the little group covariant development of the massive spinor-helicity formalism~\cite{Arkani-Hamed:2017jhn}, one of the pressing problems is to find a systematic way to construct higher point massive amplitudes to describe physical processes and with the contact term ambiguities properly addressed. 
In this paper, we have addressed this problem with two classes of momentum shift, ALT shift and mBCFW shift. 
These momentum shifts allow us to systematically construct not only the residue of the poles 
but also the (dependent) contact terms in contrast to a simple ``gluing'' of lower point amplitudes. 
After applying the momentum shifts, we have calculated four-point and five-point amplitudes for massive QED,
solving the ambiguity related to the contact terms raised in~\cite{Christensen:2022nja}.
In this paper, we have focused on massive QED, but the scope is more general, and these momentum shifts can be applied to properly construct amplitudes in other constructible theories, such as massive QCD.

We have studied the general properties of the ALT shift and mBCFW shift.
The latter relies on a specific choice of the spin axis and works only for a specific spin state of the external particles.
On the other hand, the former relies on the helicity basis and works for any combination of the spin states of the external particles.
Shifting all the external momenta allows us to study the large-$z$ behavior of the amplitude in a simple dimensional analysis,
guaranteeing the on-shell constructibility of the amplitudes of our interest without referring to a Lagrangian. 
Even though scattering amplitudes are little group covariant,
these ``good'' momentum shifts require us to specify the spin states of external particles
and thus break the little group covariance at the intermediate steps of the calculation.
It would be appealing if there exists a massive momentum shift that does not break the little group covariance, 
but this might be too ambitious,
given that it is essential to specify the helicity of external particles to define proper momentum shifts, even in the massless case.

In this paper, we have considered only the massive spin-1/2 and massless spin-1 particles, as our focus was on massive QED.
However, the ALT shift we proposed in this paper 
can be naturally extended to more general theories, such as massive QCD 
and theories with massive spin-1 particles.
Therefore, the all-line transverse shift can construct higher-point amplitudes for spontaneously broken gauge theories,
such as the electroweak sector of the SM.
This will be the main subject of our future publication~\cite{Ema:2024rss}. 
Extensive work has been done to build higher point amplitude for the electroweak sector and SMEFT from the perspective of constructing amplitude~\cite{Durieux:2019eor,Christensen:2018zcq,Bachu:2019ehv,Li:2022tec,Christensen:2024bdt,Christensen:2024xzs,Aoude:2019tzn,Liu:2022alx} without an introduction of momentum shifts. 
To better understand the on-shell constructibility of these theories at tree-level, 
one may apply our momentum shift and categorize constructible and non-constructible contact contributions to higher-point amplitudes.

The ALT shift allows us to construct higher-point amplitudes in renormalizable theories 
with up to spin-1 massless/massive particles solely from lower-point amplitudes.
This would provide a useful mathematical tool to study the general properties of these amplitudes.
Massless amplitudes in gauge theory and gravity are known to possess fascinating properties 
such as color-kinematic duality and double-copy relations~\cite{Kawai:1985xq,Bern:2008qj,HenryTye:2010tcy,Bern:2010yg,Feng:2010my,Bjerrum-Bohr:2010pnr,Mafra:2011kj,Fu:2012uy}.
It would be, therefore, interesting to see if and how the introduction of the mass affects these relations.
Another interesting direction is to extend our momentum shifts to higher spin particles.
An initial investigation indicates that the ALT shift modifies 
polarization vectors/spinors for longitudinal modes with spin higher than one. Still, a more thorough exploration is required to draw a definite conclusion.

Furthermore, to the more practical level of massive amplitudes for high-energy physics, our proposed shift might allow for unambiguous and consistent recursive relations, which might eventually help automate massive event generation.

\subsubsection*{Acknowledgement}
We acknowledge  C.~Cheung, N.~Christensen, T.~Cohen, G.~Durieux, B.~Feng, T.~Kitahara, D.~Liu, Y.~Shadmi and L.T.~Wang for helpful discussion. 
This work is supported in part by U.S. Department of Energy Grant No.~DESC0011842.  We acknowledge support from the Simons Foundation Targeted Grant 920184  to the Fine Theoretical Physics Institute, enabling a visitor program that positively impacted our work.

\appendix

\section{Massive spinor formalism}
\label{app:convention}

We will work with mostly negative metric $\eta_{\mu \nu} = \mathrm{diag}(+1,-1,-1,-1)$. 
The Lorentz and massive little group indices are raised and lowered using
\begin{align}
    \epsilon_{a b} = \epsilon_{\dot{a} \dot{b}} =  - \epsilon^{a b} = - \epsilon^{\dot{a} \dot{b}}= \begin{pmatrix}
        0  & -1 \\
        1 & 0
    \end{pmatrix},
    \quad
     \epsilon_{IJ} =  - \epsilon^{IJ}
     = \begin{pmatrix}
        0  & -1 \\
        1 & 0
    \end{pmatrix},
\end{align}
where we use the capital Latin characters to denote little group indices. 
With this convention, the little group indices are raised and lowered as\footnote{
	Several literatures, such as~\cite{Franken:2019wqr}, use a different convention that $\epsilon^{12} = \epsilon_{12} = +1$.
	With this convention, the indices are raised and lowered by 
	$\lvert \textbf{i} \rangle_{a}^I = \epsilon^{IJ} \lvert \textbf{i} \rangle_{aJ}$, 
	and
	$\lvert \textbf{i} \rangle_{aI} = \lvert \textbf{i} \rangle_{a}^J\epsilon_{JI}$.
	We do not use this convention and instead define $\epsilon_{IJ}$ as the inverse matrix of $\epsilon^{IJ}$.
}
\begin{align}
	\lvert \textbf{i} \rangle_{a}^I = \epsilon^{IJ}	\lvert \textbf{i} \rangle_{aJ},
	\quad
	\lvert \textbf{i} \rangle_{aI} = \epsilon_{IJ}\lvert \textbf{i} \rangle_{a}^J.
\end{align}
Then, we write the momentum for a massive particle in the spinor space as
\begin{align}
    p_{a \dot{a}} = p_\mu [\sigma^{\mu}]_{a\dot{a}} 
    = \epsilon^{JI} \lvert \textbf{i} \rangle_{aI} [ \textbf{i} \lvert_{\dot{a}J} ,
    \quad
    p^{ \dot{a} a} = p_\mu [\bar{\sigma}^{\mu}]^{\dot{a}a}
    = \epsilon_{IJ} \lvert \textbf{i}]^{\dot{a}J} \langle \textbf{i} \lvert^{aI},
\end{align}
where $\sigma^\mu = (\mathbbm{1}, \vec{\sigma})$ and $\bar{\sigma}^{\mu} = (\mathbbm{1}, -\vec{\sigma})$
with $\sigma^i$ being the Pauli matrix.
We note that it satisfies $\epsilon^{ab}\epsilon^{\dot{a}\dot{b}}[\sigma^\mu]_{b\dot{b}} = [\bar{\sigma}^\mu]^{\dot{a}a}$.
The on-shell condition $p^2 = m_i^2$ translates to
\begin{align}
	\mathrm{det}[p_{a\dot{a}}] = \mathrm{det}[\vert\mathbf{i}\rangle_{a}^{I}] 
	\times \mathrm{det}[[\mathbf{i}\vert_{\dot{a}I}] = m_i^2,
\end{align}
and we fix the normalization of the massive spinors as
\begin{align}
	\mathrm{det}[\vert \mathbf{i}\rangle_{a}^{I}] 
	= \mathrm{det}[[\mathbf{i}\vert_{\dot{a}I}] = m_i.
\end{align}
With this normalization, we have
\begin{align}
	\langle \mathbf{i}^{I}\mathbf{i}^{J}\rangle = -\epsilon^{IJ}m_i,
	\quad
	[\mathbf{i}_I \mathbf{i}_J] = -\epsilon_{IJ} m_i,
	\label{eq:massive_spinor_products}
\end{align}
where the products of the spinors are defined as
\begin{align}
	\langle i j \rangle = \langle i\vert^{a} \vert j\rangle_{a} = \epsilon^{ab} \vert j\rangle_a \vert i \rangle_b,
	\quad
	[i j] = [i\vert_{\dot{a}} \vert j ]^{\dot{a}} = \epsilon_{\dot{a}\dot{b}} \vert j]^{\dot{a}}\vert i]^{\dot{b}}.
\end{align}
If we flip the sign of the momentum, $p_i \rightarrow -p_i$, we follow the convention
\begin{align}
    \lvert -\textbf{i} \rangle = \lvert \textbf{i} \rangle, \quad    \lvert -\textbf{i} ]= -\lvert \textbf{i} ].
\end{align}
Finally, momentum for a massless particle satisfies $\mathrm{det}[p_{a\dot{a}}] = 0$ and hence
can be expressed as
\begin{align}
	p_{a\dot{a}} = \vert i \rangle_a [i\vert_{\dot{a}},
\end{align}
by only a single pair of spinors.
With Eq.~\eqref{eq:massive_spinor_products}, we can show that the massive spinors satisfy
\begin{align}
     p^{ \dot{a} a} \lvert \textbf{i} \rangle^I_a = m_i \lvert \textbf{i} ]^{\dot{a}I}, 
     \quad
     p_{  a \dot{a}} \lvert \textbf{i} ]^{\dot{a}I} = m_i \lvert \textbf{i} \rangle^{I}_a,
     \quad
      [\textbf{i} \lvert^{I}_{\dot{a}}p^{ \dot{a} a}  = - m_i \langle \textbf{i} ]^{aI}, 
      \quad
      \langle \textbf{i} \lvert^{{a}I} p_{  a \dot{a}}= - m_i [ \textbf{i} \lvert_{\dot{a}}^{I},
      \label{eq:Dirac_eq}
\end{align}
which are nothing but the Dirac equations.
Therefore, Dirac spinors are written as
\begin{align}
    u^I(p) = \begin{pmatrix}
        	\lvert \textbf{i} \rangle_{a}^I \\ 
        \lvert \textbf{i} ]^{\dot{a}I}
    \end{pmatrix},  \quad   v^I(p) = \begin{pmatrix}
        	\lvert \textbf{i} \rangle_{a}^I \\ 
       - \lvert \textbf{i} ]^{\dot{a}I}
    \end{pmatrix},
    \quad
    \bar{u}^I(p) = \begin{pmatrix}
        - \langle \textbf{i}  \lvert^{aI} & [\textbf{i}  \lvert_{\dot{a}}^I  
    \end{pmatrix},\quad  \bar{v}^I(p) = \begin{pmatrix}
         \langle \textbf{i}  \lvert^{aI} & [\textbf{i}  \lvert_{\dot{a}}^I
        \end{pmatrix},
        \label{eq:fermion_wavefn}
\end{align}
and they satisfy
\begin{align}
	\epsilon_{IJ}u^I(p)\bar{u}^J(p) = \slashed{p} + m,
	\quad
	\epsilon_{IJ}v^{I}(p)\bar{v}^{J}(p) = \slashed{p}-m.
\end{align}

\subsubsection*{Helicity basis}

So far, the spin quantization axis has been taken arbitrarily.
In the main text, we rely on the helicity basis when using the all-line transverse shift.
In the helicity basis, we expand $\lvert\textbf{i}\rangle_a^I$ and $[\textbf{i}\lvert_{\dot{a}}^I$ as
\begin{align}
   \lvert \textbf{i} \rangle^I_a = \lvert i \rangle_a \delta^I_- + \lvert \eta_i \rangle_a \delta^{I}_+, \quad [ \textbf{i} \lvert^I_{\dot{a}} = [ i \lvert_{\dot{a}} \delta_+^I + [\eta_i \lvert _{\dot{a}} \delta^{I}_-,
\end{align}
When using this basis, we use $I=\pm$ to indicate little group indices, and so
$\vert \mathbf{i}\rangle^{+} = \vert \eta_i \rangle,\vert \mathbf{i}\rangle^{-} = 
\vert i\rangle,[\mathbf{i}\vert^{+} = [i\vert,[\mathbf{i}\vert^{-} = [\eta_{i}\vert$. 
The explicit form of these spinors is given by
\begin{align}
	&\vert i\rangle_a = \sqrt{E_i+p_i}\begin{pmatrix} -s_i^* \\ c_i \end{pmatrix},
	\quad
	[i\vert_{\dot{a}} = \sqrt{E_i + p_i}\begin{pmatrix} -s_i & c_i \end{pmatrix},
	\\
	&\vert \eta_i \rangle_a = \sqrt{E_i-p_i}\begin{pmatrix} c_i \\ s_i \end{pmatrix},
	\quad
	[\eta_i\vert_{\dot{a}} = - \sqrt{E_i-p_i}\begin{pmatrix} c_i & s_i^* \end{pmatrix}.
\end{align}
where $c_i = \cos \frac{\theta_i}{2}, s_i=e^{i\phi_i} \sin \frac{\theta_i}{2}$
with $\theta_i$ and $\phi_i$ the solid angles of the momentum.
The spinor products are given by
\begin{align}
	\langle i \eta_i \rangle = [i\eta_i ] = m_i.
\end{align}
In the massless limit, we can take $\langle i \eta_i \rangle \rightarrow 0$ and $[ i \eta_i ] \rightarrow 0$ in the complex plane to find
\begin{align}
    \lvert \textbf{i} \rangle^- \rightarrow \lvert i \rangle, 
    \quad 
    [ \textbf{i}\lvert^+ \rightarrow [ i \lvert,
    \quad
    \frac{\lvert \textbf{i}\rangle^+}{m_i} \rightarrow \frac{\lvert \xi \rangle}{\langle  i \xi  \rangle},
    \quad
    \frac{[ \textbf{i}\lvert^-}{m_i} \rightarrow \frac{[ \xi \lvert}{[  i \xi ]}, 
    \label{eq: massless_limit}
\end{align}
where $\xi$ is arbitrary spinor satisfying $ \langle i \xi \rangle = [i \xi] \neq 0$. 
For massive momentum $p_i$, polarization vectors are given by
\begin{align}
	\epsilon_i^{(+)} = \sqrt{2}\frac{\vert \eta_i\rangle [ i \vert}{m_i},
	\quad
	\epsilon_i^{(-)} = -\sqrt{2}\frac{\vert i\rangle [ \eta_i \vert}{m_i},
	\quad
	\epsilon_i^{(L)} = \frac{\vert i\rangle [i \vert + \vert \eta_i\rangle [\eta_i \vert}{m_i}.
	\label{eq:massive_pol_vectors_app}
\end{align}
where ``$\pm$" corresponds to the transverse mode while ``$L$" denotes the longitudinal mode.
Thus we see that $\langle i \vert \gamma^0 \vert \eta_i] = \langle \eta_i \vert \gamma^0 \vert i] = 0$ 
is a statement that the zeroth component of the transverse polarization vector vanishes. 
By taking the massless limit of the transverse modes, polarization vectors for massless momentum $p_i$ are expressed as
\begin{align}
	\epsilon_i^{(+)} = \sqrt{2}\frac{\vert \xi_i\rangle [ i \vert}{\langle i \xi_i\rangle},
	\quad
	\epsilon_i^{(-)} = -\sqrt{2}\frac{\vert i\rangle [ \xi_i \vert}{[i\xi_i]}.
\end{align}

\section{Computational details}
\label{app:details}

In this appendix, we provide computational details that we omit in the main text.

\subsection{Four-point amplitude}
\label{app:details_4pt}
Let us start from Eq.~\eqref{eq:four-point-pole}
\begin{align}
	\sum_{\lambda}\hat{A}_3(\mathbf{12}I^\lambda) \times \hat{A}_3(\mathbf{34}{-I}^{-\lambda})
	&= \tilde{e}^2\left(\frac{\langle \xi \hat{\textbf{2}} \rangle [ \hat{\textbf{1}} \hat{I}]}{ \langle \xi \hat{I}\rangle} + \frac{\langle  \hat{\textbf{1}} \xi \rangle [\hat{I} \hat{\textbf{2}} ]}{\langle \xi \hat{I}\rangle} \right) \left( \frac{[ \xi {\hat{\textbf{4}}} ] \langle \hat{\textbf{3}} -\hat{I} \rangle}{ [ \xi -\hat{I} ] } + \frac{[  \hat{\textbf{3}} \xi ] \langle -\hat{I} {\hat{\textbf{4}}} \rangle}{[ \xi -\hat{I} ] } \right ) + \left([...] \leftrightarrow \langle ... \rangle\right), \label{eq:app_c1}
\end{align}
We will explicitly work out two of the terms
\begin{align}
	    \frac{\langle \xi  \hat{\textbf{2}} \rangle [ \hat{\textbf{1}} I]}{ \langle \xi \hat{I}\rangle}  \frac{[ \xi  \hat{\textbf{4}} ] \langle \hat{\textbf{3}} -\hat{I} \rangle}{ [ \xi  -\hat{I} ] } +  \frac{[  \hat{\textbf{1}} \xi  ] \langle \hat{I} \hat{\textbf{2}} \rangle}{[ \xi  \hat{I} ] } \frac{\langle  \hat{\textbf{3}} \xi  \rangle [-\hat{I} \hat{\textbf{4}} ]}{\langle \xi  -\hat{I}\rangle}  =
           \frac{[\hat{\textbf{1}} \lvert \hat{p}_I \lvert \hat{\textbf{3}} \rangle  [\hat{\textbf{4}} \lvert p_\xi  \lvert \hat{\textbf{2}} \rangle}{ 2 \hat{p}_I . p_\xi  }   +  \frac{[\hat{\textbf{1}} \lvert p_\xi  \lvert \hat{\textbf{3}} \rangle  [\hat{\textbf{4}} \lvert \hat{p}_I \lvert \hat{\textbf{2}} \rangle }{2 \hat{p}_I . p_\xi  }.
\end{align}
Then we apply the following identity~\cite{Christensen:2019mch}
\begin{equation}
    \langle i \lvert p_m \lvert j ]     \langle k \lvert p_n \lvert l ] \ = \ -[lj] \langle k \lvert p_n p_m \lvert i \rangle  +  [ l \lvert p_m \lvert i \rangle  \langle k \lvert p_n \lvert j], \label{eq:schouten_momentum}
\end{equation}
we can simplify our expression
\begin{align}\label{eq:exp_2}
  &  - \frac{[\hat{\textbf{4}}\hat{\textbf{1}}] \langle \hat{\textbf{2}} \lvert p_\xi  \hat{p}_I \lvert \hat{\textbf{3}}\rangle}{2 \hat{p}_I \cdot p_\xi } + \frac{[ \hat{\textbf{4}} \lvert \hat{p}_I \lvert \hat{\textbf{3}} \rangle \langle \hat{\textbf{2}} \lvert p_\xi  \lvert \hat{\textbf{1}} ]}{2 \hat{p}_I \cdot p_\xi }  
   - \frac{[\hat{\textbf{4}}\hat{\textbf{1}}] \langle \hat{\textbf{2}} \lvert \hat{p}_I p_\xi  \lvert \hat{\textbf{3}}\rangle}{2 \hat{p}_I . p_\xi } + \frac{[ \hat{\textbf{4}} \lvert p_\xi  \lvert \hat{\textbf{3}} \rangle \langle \hat{\textbf{2}} \lvert \hat{p}_I \lvert \hat{\textbf{1}} ]}{2 \hat{p}_I \cdot p_\xi } \nonumber \\ 
    =& - \frac{[\hat{\textbf{4}}\hat{\textbf{1}}] \langle \hat{\textbf{2}} \lvert p_\xi  \hat{p}_I + \hat{p}_I p_\xi  \lvert \hat{\textbf{3}}\rangle}{2 \hat{p}_I \cdot p_\xi }  + \frac{[ \hat{\textbf{4}} \lvert \hat{p}_I \lvert \hat{\textbf{3}} \rangle \langle \hat{\textbf{2}} \lvert p_\xi  \lvert \hat{\textbf{1}} ]}{2 \hat{p}_I \cdot p_\xi }    
    +\frac{[ \hat{\textbf{4}} \lvert p_\xi  \lvert \hat{\textbf{3}} \rangle \langle \hat{\textbf{2}} \lvert \hat{p}_I \lvert \hat{\textbf{1}} ]}{2 \hat{p}_I \cdot p_\xi }.
\end{align}
The last two terms will cancel from other expressions of the form $\textbf{1} \leftrightarrow \textbf{2}$ and $\textbf{3} \leftrightarrow \textbf{4}$ in Eq.~\eqref{eq:app_c1} since $\langle\textbf{i}\lvert p_{ij}\lvert \textbf{j}] = - \langle\textbf{j}\lvert p_{ij}\lvert \textbf{i}]$. After using gamma matrix identity $\{ \gamma_{\mu}, \gamma_{\nu}\}=2\eta_{\mu \nu}$, the first term simplifies to $\langle\hat{\textbf{2}}\hat{\textbf{3}} \rangle [\hat{\textbf{1}}\hat{\textbf{4}}]$. Following similar procedure for rest of the terms in Eq.~\eqref{eq:app_c1}, we obtain the expression in Eq.~\eqref{eq:A3prod_unhat}
\begin{align}
	\sum_{\lambda}\hat{A}_3(\mathbf{12}I^\lambda) \times \hat{A}_3(\mathbf{34}{-I}^{-\lambda})
	= \tilde{e}^2\left[
	\langle {\hat{\textbf{1}}} {\hat{\textbf{3}}} \rangle [{\hat{\textbf{2}}}{\hat{\textbf{4}}}] + \langle {\hat{\textbf{1}}}{\hat{\textbf{4}} }\rangle
	[{\hat{\textbf{2}}}{\hat{\textbf{3}}}] +\langle {\hat{\textbf{2}}}{\hat{\textbf{3}}} \rangle 
	[{\hat{\textbf{1}}}{\hat{\textbf{4}}}] +\langle {\hat{\textbf{2}}} {\hat{\textbf{4}} }\rangle [{\hat{\textbf{1}}}{\hat{\textbf{3}}}]  \right].
\end{align}
Next, we show another derivation of the four-point amplitude using $[ \textbf{1}\textbf{3} \rangle$ mBCFW shift. A good shift exists for spin projection $(s_1,s_3)= (1,2)$, and the other particles can have any spin. The shift is of the following form
\begin{align}
    [\hat{\textbf{1}}\lvert^2 &= [l_1\lvert -z[l_3\lvert,
    \quad
    \lvert \hat{\textbf{3}} \rangle^1 = \lvert l_3 \rangle + z \lvert l_1 \rangle.
\end{align}
Then, for the shift, we find
\begin{align}
       \frac{\hat{x}_{12}} {\hat{x}_{34}}  
     &=  -  \frac{m_\mu}{m_e}\frac{\langle \xi \lvert p_2 \hat{p}_I \lvert \xi \rangle }{ \langle \xi \lvert {p_4} \hat{p}_I \lvert \xi \rangle} 
          =    \frac{m_\mu m_e}{\langle l_1 l_3 \rangle [l_3 l_1]}\frac{\langle l_1 \lvert p_2 \lvert l_3] \langle l_3 l_1 \rangle }{ \langle l_1 \lvert {p_4} \lvert l_3] \langle l_3 l_1 \rangle}
                    =   - \frac{m_\mu m_e}{\langle l_1 l_3 \rangle [l_3 l_1]},
    \\
    \frac{\hat{{x}}_{34}}{\hat{{x}}_{12}}&= - \frac{\langle l_1 l_3 \rangle [l_3 l_1]}{m_{e}m_{\mu}}.
\end{align}
Here we used $\lvert \xi \rangle = \lvert l_1 \rangle$ and using conservation of momentum to get $\langle l_1 \lvert p_2 \lvert l_3] =-\langle l_1 \lvert p_4 \lvert l_3] $. Now we can use Eq.~\eqref{3pt}, and sum over helicities at the factorization channel
 \begin{align}
    A_4 &= \frac{\tilde{e}^2}{s} \bigg(\frac{\hat{x}_{12}} {\hat{x}_{34}} \langle \hat{\textbf{1}}^1 \textbf{2} \rangle [\hat{\textbf{3}}^2\textbf{4}] + \frac{\hat{x}_{34}} {\hat{x}_{12}} \langle \hat{\textbf{3}}^2\textbf{4} \rangle [\hat{\textbf{1}}^1 \textbf{2}] \bigg) \nonumber \\
   &= \frac{\tilde{e}^2}{s} \bigg(  \frac{m_\mu m_e}{\langle l_1 l_3 \rangle [l_3 l_1]} \langle l_1 \textbf{2} \rangle [l_3 \textbf{4}] 
   - \frac{\langle l_1 l_3 \rangle [l_3 l_1]}{m_\mu m_e} \frac{m_\mu}{\langle l_3 l_1 \rangle} \langle l_1 \textbf{4} \rangle  \frac{m_e}{[l_3 l_1]}[l_3 \textbf{2}]\bigg)
   =  \frac{\tilde{e}^2}{s} \big( \langle \textbf{2} \textbf{3}^2 \rangle [\textbf{1}^1 \textbf{4}] + \langle \textbf{1}^1 \textbf{4} \rangle [\textbf{2} \textbf{3}^2] \big).
\end{align} 
Here terms of the form $\langle \textbf{1}\textbf{3} \rangle [\textbf{2}\textbf{4}]$ and $[ \textbf{1}\textbf{3} ] \langle \textbf{2}\textbf{4} \rangle$ do not appear since $\langle \textbf{1}^1\textbf{3}^2 \rangle= [ \textbf{1}^1\textbf{3}^2 ]=0 $ due to the choice of spin axis for particle 1 and 3 required by BCFW shift. In general, using massive BCFW-type shifts where a ``good" does not exist for all spin projections, we need to resort to the spin raising and lowering operator to construct the amplitude for all spin projections, which might not be easily realizable for more complex theories.

Now we show the computation details from Eq.~\eqref{christensen} to Eq.~\eqref{christensen3}. Note that we did not use the on-shell condition of the internal photon during the derivation, so we will omit the hats here.
We start by simplifying $\ssss{\bn{12}}\ssss{\bn{3|21|4}}$:
\begin{align}
    \ssss{\bn{12}}\ssss{\bn{3|21|4}}&=\ssss{\bn{24}}[\bn{1|12|3}]-[\bn{2|12|3}][\bn{14}] \nonumber\\
    &=-m_e\bn{[24]\aass{1|2|3}+[2|21|3][14]}-2p_1\cdot p_2\bn{[23][14]} \nonumber\\
    &=-m_e\left(\bn{[24]\aass{1|2|3}+[14]\aass{2|1|3}}\right)-2p_1\cdot p_2\bn{[23][14]},
    \label{chain1}
\end{align}
where in the first step we used generalized Schouten identities\cite{Christensen:2019mch}, in the second step we used equations of motion and $\bn{12+21}=2p_1\cdot p_2$, and in the last step we used equations of motion. Using momentum conservation and applying generalized Schouten identities one more time with equations of motion, we get
\begin{align}
    \bn{[24]\aass{1|2|3}}&=-\bn{[24]\aass{1|1+3+4|3}}\nonumber\\
    &=\bn{[24]}\left(m_1\bn{[13]}-m_3\bn{\aaaa{13}}\right)-\bn{[24]\aass{1|4|3}}\nonumber\\
    &=\bn{[24]}\left(m_e\bn{[13]}-m_\mu\bn{\aaaa{13}}\right)+\bn{[43]\aass{1|4|2}+[32]\aass{1|4|4}}\nonumber\\
    &=\bn{[24]}\left(m_e\bn{[13]}-m_\mu\bn{\aaaa{13}}\right)-\frac{1}{m_e}\bn{[43][1|14|2]}+m_\mu\bn{[32]\aaaa{14}}\nonumber\\
    &=m_e\bn{[13][24]}-m_\mu\left(\bn{\aaaa{14}[23]+\aaaa{13}[24]}\right)-\frac{1}{m_e}\bn{[43][1|14|2]},
\end{align}
and a similar equation with the role of $\bn{1}$ and $\bn{2}$ interchanged
\begin{align}
    \bn{[14]\aass{2|1|3}}=m_e\bn{[23][14]}-m_\mu\left(\bn{\aaaa{24}[13]+\aaaa{23}[14]}\right)-\frac{1}{m_e}\bn{[43][2|24|1]}.
\end{align}
We reverse the spinor chain $\bn{[2|24|1]}$ and apply momentum conservation to get
\begin{align}
    \bn{[43][2|24|1]}=-\bn{[43][1|42|2]}=\bn{[43][1|4(1+3+4)|2]}=\bn{[43]}\left(\bn{[1|43|2]+[1|41|2]}\right)- m_\mu^2\bn{[12][34]},
\end{align}
which then gives
\begin{align}
    \bn{[43][1|14|2]+[43][2|24|1]=[43][1|43|2]}-2p_1\cdot p_4 \bn{[12][34]}-m_\mu^2\bn{[12][34]},
\end{align}
and
\begin{align}
    \bn{[24]\aass{1|2|3}+[14]\aass{2|1|3}}=&m_e\left(\bn{[13][24]+[23][14]}\right)-m_\mu\left(\bn{\aaaa{14}[23]+\aaaa{13}[24]+\aaaa{24}[13]+\aaaa{23}[14]}\right)\nonumber\\
    &-\frac{1}{m_e}\left(\bn{[43][1|43|2]}-2p_1\cdot p_4\bn{[12][34]}-m_\mu^2\bn{[12][34]}\right),
\end{align}
substitute this back to Eq.~\eqref{chain1}, we get
\begin{align}
    \ssss{\bn{12}}\ssss{\bn{3|21|4}}=&-m_e^2\left(\bn{[13][24]+[23][14]}\right)+m_e m_\mu \left(\bn{\aaaa{14}[23]+\aaaa{13}[24]+\aaaa{24}[13]+\aaaa{23}[14]}\right)\nonumber\\
    &-2p_1\cdot p_4 \bn{[12][34]}-2p_1\cdot p_2 \bn{[23][14]}-m_\mu^2\bn{[12][34]}+\bn{[43][1|43|2]}.
\end{align}
Comparing with Eq.~\eqref{christensen}, we see the last term in the equation above cancels with $\bn{[34][1|43|2]}$ in Eq.~\eqref{christensen}, this allows us to write the final result in a form where no momentum is sandwiched between the spinor products. Using Schouten identity and definition of Mandelstam variables, we get
\begin{align}
    \bn{[12][3|21|4]+[34][1|43|2]}=m_em_\mu(\bn{[13]\aaaa{24}+[24]\aaaa{13}+[14]\aaaa{23}+[23]\aaaa{14}})-\bn{s[14][23]-u[12][34]},
\end{align}
which leads to Eq.~\eqref{christensen3} upon plugging into Eq.~\eqref{christensen}.

\subsection{Five-point amplitude}
\label{app:details_5pt}

To calculate the five-point amplitude, we would require Compton scattering results. 
\begin{align}
     &   A_4(\textbf{1},\textbf{2},3^-,4^-)  =  -\frac{\tilde{e}^2m_f \langle {3}4\rangle^2 [\textbf{1}\textbf{2}]  }{(p_{13}^2 - m_f^2) (p_{14}^2 - m_f^2)}, \\
     &   A_4(\textbf{1},\textbf{2},3^+,4^+)  =  -\frac{\tilde{e}^2 m_f [ {3}4 ]^2 \langle \textbf{1}\textbf{2} \rangle  }{(p_{13}^2 - m_f^2) (p_{14}^2 - m_f^2)}, \\
     &    A_4(\textbf{1},\textbf{2},3^-,4^+) =  -\frac{ \tilde{e}^2\langle 3 \lvert p_1 - p_2\lvert 4 ] (\langle \textbf{2} 3 \rangle [\textbf{1} 4] + \langle \textbf{1} 3 \rangle [\textbf{2} 4]) }{2(p_{13}^2 - m_f^2) (p_{14}^2 - m_f^2)}, \\
          &   A_4(\textbf{1},\textbf{2},3^+,4^-) =  -\frac{ \tilde{e}^2\langle 4 \lvert p_1 - p_2 \lvert 3 ] (\langle \textbf{2} 4 \rangle [\textbf{1} 3] + \langle \textbf{1} 4 \rangle [\textbf{2} 3]) }{2(p_{13}^2 - m_f^2) (p_{14}^2 - m_f^2)},
\end{align}
where $m_f=m_1=m_2$. We will show the calculation for negative helicity photon. For $\hat{p}_{35}^2=m_{\mu}^2$ pole, we have
\begin{align}
\sum_{\lambda}\hat{A}_4(\mathbf{124}I^\lambda) \times \hat{A}_3(\mathbf{3}5^-{-I}^{-\lambda})\Big|_{\hat{p}_{35}^2 = m_\mu^2} =&\frac{\tilde{e}^3}{{p}_{12}^2}\left(
	\langle {{\hat{\textbf{1}}}} {{\hat{\textbf{I}}}} \rangle [{{\hat{\textbf{2}}}}{{\hat{\textbf{4}}}}] + \langle {{\hat{\textbf{1}}}}{{\hat{\textbf{4}}}}\rangle
	[{{\hat{\textbf{2}}}}{{\hat{\textbf{I}}}}] +\langle {{\hat{\textbf{2}}}}{{\hat{\textbf{I}}}} \rangle 
	[{{\hat{\textbf{1}}}}{{\hat{\textbf{4}}}}] +\langle {{\hat{\textbf{2}}}} {{\hat{\textbf{4}} }}\rangle [{{\hat{\textbf{1}}}}{{\hat{\textbf{I}}}}]  \right) \hat{\tilde{{x}}}_{I3}[-{\hat{\textbf{I}}}{\hat{\textbf{3}}}].
 \end{align}
 Then, working with two of the terms
 \begin{align}
    (\langle \hat{\textbf{1}} \hat{\textbf{4}} \rangle [\hat{\textbf{2}} I]^X + [\hat{\textbf{1}}\hat{\textbf{4}}] \langle \hat{\textbf{2}}I^X\rangle)\hat{\tilde{{x}}}_{I3} [-\hat{\textbf{I}}_X \hat{\textbf{3}}] 
  &= - \hat{\tilde{{x}}}_{I3} m_\mu \langle \hat{\textbf{1}} \hat{\textbf{4}} \rangle [\hat{\textbf{2}}\hat{\textbf{3}}] - \hat{\tilde{{x}}}_{I3} [\hat{\textbf{1}}\hat{\textbf{4}}] \langle \hat{\textbf{2}} \lvert \hat{p}_I \lvert \hat{\textbf{3}}]   \nonumber \\
    &= -\hat{\tilde{{x}}}_{I3} m_\mu (\langle \hat{\textbf{1}} \hat{\textbf{4}} \rangle [\hat{\textbf{2}}\hat{\textbf{3}}] + [\hat{\textbf{1}}\hat{\textbf{4}}] \langle \hat{\textbf{2}} \hat{\textbf{3}} \rangle) + [\hat{\textbf{1}}\hat{\textbf{4}}] \langle \hat{\textbf{2}}\hat{5} \rangle [\hat{5} \hat{\textbf{3}}] [q \lvert \frac{\hat{p}_3}{m_\mu} \lvert \hat{5} \rangle/[q\hat{5}] \nonumber \\
       &= - \hat{\tilde{{x}}}_{I3} m_\mu (\langle \hat{\textbf{1}} \hat{\textbf{4}} \rangle [\hat{\textbf{2}}\hat{\textbf{3}}] + [\hat{\textbf{1}}\hat{\textbf{4}}] \langle \hat{\textbf{2}} \hat{\textbf{3}} \rangle) + [\hat{\textbf{1}}\hat{\textbf{4}}] \langle \hat{\textbf{2}}\hat{5} \rangle \langle \hat{\textbf{3}} \hat{5} \rangle.
\end{align}
In the first equality we used $\lvert \hat{\textbf{I}}^X \rangle [ - \hat{\textbf{I}}_X\lvert =-\hat{p}_I$ and $\lvert \hat{\textbf{I}}^X ] [ - \hat{\textbf{I}}_X\lvert =-m_I$.  For the second equality we substituted $\hat{p}_I = \hat{p}_3 + \hat{p}_5$; for the last line we applied Schouten identity and $\hat{p}_3 \cdot \hat{p}_5=0$. Now, we can write 
\begin{align}
    -m_{\mu}\hat{\tilde{x}}_{I3}=  \frac{\langle \hat{5} \lvert \hat {p}_3 \lvert q]}{[q \hat{5} ]}  =     \frac{\langle \hat{5} \lvert \hat {p}_3 \lvert q] 2 \hat{p}_4 \cdot \hat{p}_5}{[q \hat{5} ] 2 \hat{p}_4 \cdot \hat{p}_5} 
     =   \frac{\langle 5 \lvert \hat {p}_3 \lvert q] \langle 5 \lvert \hat{p}_4 \lvert \hat{5}]}{[q \hat{5} ] 2 \hat{p}_4 \cdot \hat{p}_5} 
     =  \frac{\langle 5 \lvert \hat {p}_4 \hat{p}_3 \lvert 5 \rangle}{2 \hat{p}_4 \cdot \hat{p}_5}.
\end{align}
In the last equality we used Eq.~\eqref{eq:schouten_momentum} and $\hat{p}_3\cdot\hat{p}_5=0$. Thus we derive Eq.~\eqref{eq:p3p5}
 \begin{align}
\sum_{\lambda}\hat{A}_4(\mathbf{124}I^\lambda) \times \hat{A}_3(\mathbf{3}5^-{-I}^{-\lambda})\Big|_{\hat{p}_{35}^2 = m_\mu^2} 
 =& \frac{\tilde{e}^3}{\hat{p}_{12}^2} \bigg[ \frac{\langle \hat{5} \lvert \hat{p}_4 \hat{p}_3 \lvert \hat{5} \rangle }{2\hat{p}_4 \cdot \hat{p}_5} \left( 	\langle {\hat{\textbf{1}}} \hat{{\textbf{3}}} \rangle [\hat{{\textbf{2}}}\hat{{\textbf{4}}}] + \langle \hat{{\textbf{1}}}\hat{{\textbf{4}}}\rangle
	[\hat{{\textbf{2}}}\hat{{\textbf{3}}}] +\langle \hat{{\textbf{2}}}\hat{{\textbf{3}}} \rangle 
	[\hat{{\textbf{1}}}\hat{{\textbf{4}}}] +\langle \hat{{\textbf{2}}} \hat{{\textbf{4} }}\rangle [\hat{{\textbf{1}}}\hat{{\textbf{3}}}] \right) \nonumber \\
& \qquad  + [\hat{\textbf{1}} \hat{\textbf{4}}] \langle \hat{\textbf{2}} \hat{ 5} \rangle \langle \hat{\textbf{3}} \hat{5} \rangle + [\hat{\textbf{2}}\hat{\textbf{4}}] \langle \hat{\textbf{1}} \hat{5} \rangle \langle \hat{\textbf{3}} \hat{5} \rangle\bigg].
\label{eq:app_p3p5}
 \end{align}
 Similarly, we can derive Eq.~\eqref{eq:p4p5} by using $\textbf{3}\leftrightarrow\textbf{4}$.

\renewcommand{\IsBoldNumber}[1]{%
	\IfSubStr{1234}{#1}{\hat{\mathbf{#1}}}{\IfSubStr{5Ip}{#1}{\hat{#1}}{#1}}%
}

For the $\hat{p}_{12}$ pole, we have
\begin{align}
    \sum_{\lambda}\hat{A}_3(\mathbf{12}I^\lambda) \times \hat{A}_4(\mathbf{34}5^-{-I}^{-\lambda}) \Big|_{\hat{p}_{12}^2=0}=&-\tilde{e}^3\hat{x}_{12}\bn{\aaaa{12}}\frac{m_\mu\bn{[34]\aaaa{5-I}}^2}{\left(2\bn{p}_3\cdot \bn{p}_5\right)\left(2 \bn{p}_4\cdot \bn{p}_5\right)}\nonumber\\
    &-\tilde{e}^3\hat{\tilde{x}}_{12}\bn{[12]}\frac{\left(\bn{[3-I]\aaaa{45}+\aaaa{35}[4-I]}\right)\bn{\ssaa{-I|3|5}}}{\left(2\bn{p}_3\cdot \bn{p}_5\right)\left(2\bn{p}_4\cdot \bn{p}_5\right)}.
    \label{12pole}
\end{align}
We start from the first line. The x-factor can be absorbed using its definition: 
\begin{align}
    \hat{x}_{12}\bn{\aaaa{12}}\bn{[34]\aaaa{5-I}}^2=\hat{x}_{12}\bn{\aaaa{12}}\bn{[34]\aaaa{5I}}^2=\bn{\aaaa{12}}\bn{[34]}\aaaa{\bn{5}|\hat{x}_{12}|\hat{I}}\aaaa{\bn{5}\bn{I}}=\frac{1}{m_e}\bn{\aaaa{12}}\bn{[34]}\aass{\bn{5|p}_{1}|\bn{I}}\aaaa{\bn{5}\bn{I}},
\end{align}
where we used the on-shell condition of the photon to rewrite $\frac{\hat{p}_1-\hat{p}_2}{2m_e}|\hat{I}\rangle$ as $\frac{\hat{p}_1}{m_e}|\hat{I}\rangle$, as we have seen in our main text, this will not cause any problems as long as we are careful with the momentum shift. Now we use $|\hat{I}]\langle \hat{I}|=\hat{I}$ and use $\aaaa{\bn{55}}=0$
\begin{align}
    \hat{x}_{12}\bn{\aaaa{12}}\bn{[34]\aaaa{5-I}}^2=-\frac{1}{m_e}\bn{\aaaa{12}}\bn{[34]}\aaaa{\bn{5|p}_1\bn{I}|\bn{5}}=\frac{1}{m_e}\bn{\aaaa{12}}\bn{[34]}\aaaa{\bn{5|1(1+2)|5}}=\frac{1}{m_e}\bn{\aaaa{12}}\bn{[34]}\aaaa{\bn{5|12|5}}.
\end{align}
Further simplifications lead to
\begin{align}
    \bn{\aaaa{12}}\bn{[34]}\aaaa{\bn{5|12|5}}&=\bn{[34]}\left(\bn{\aaaa{2|12|5}\aaaa{15}-\aaaa{25}\aaaa{1|12|5}}\right)\nonumber\\
    &=\bn{[34]}\left(2\bn{p}_1\cdot\bn{p}_2 \bn{\aaaa{25}\aaaa{15}}-\bn{\aaaa{2|21|5}\aaaa{15}}+m_e \bn{\aaaa{25}\ssaa{1|2|5}}\right)\nonumber\\
    &=\bn{[34]}\left(2p_1\cdot p_2\bn{\aaaa{25}\aaaa{15}}+m_e \bn{\ssaa{2|1|5}\aaaa{15}}+m_e \bn{\aaaa{25}\ssaa{1|2|5}}\right),
\end{align}
Momentum conservation gives
\begin{align}
    \bn{[34]}\bn{\ssaa{2|1|5}\aaaa{15}}=\bn{\aaaa{15}[34]}\left(m_e\bn{\aaaa{25}}-\bn{\ssaa{2|3|5}-\ssaa{2|4|5}}\right).
\end{align}
Using generalized Schouten identity and equations of motion, we have
\begin{align}
    &\bn{[34]\ssaa{5|3|2}}=-m_\mu\bn{[24]\aaaa{35}}-\bn{[23]\aass{5|3|4}},\\
    &\bn{[34]\ssaa{5|4|2}}=m_\mu\bn{[23]\aaaa{45}}+\bn{[24]\aass{5|4|3}}.
\end{align}
Putting everything together and using $2\hat{p}_1\cdot\hat{p}_2=-2m_e^2$, the first line of Eq.~\eqref{12pole} can be rewritten as
\begin{align}
    -\tilde{e}^3\hat{x}_{12}\bn{\aaaa{12}}\frac{m_\mu\bn{[34]\aaaa{5-I}}^2}{\left(2\bn{p}_3\cdot \bn{p}_5\right)\left(2 \bn{p}_4\cdot \bn{p}_5\right)}=&\frac{m_\mu\tilde{e}^3}{\left(2\bn{p}_3\cdot \bn{p}_5\right)\left(2 \bn{p}_4\cdot \bn{p}_5\right)}\left(m_\mu\bn{\aaaa{15}\aaaa{45}[23]}-\bn{\aaaa{15}\aass{5|3|4}[23]}\right)  \nonumber\\
    &+\bn{1\leftrightarrow 2-3\leftrightarrow 4-(1,3)\leftrightarrow(2,4)}.
    \label{firstline}
\end{align}

Now, we move on to the second line of Eq.~\eqref{12pole}. Using
\begin{align}
    \hat{\tilde{x}}_{12}|-\bn{I}]=-\hat{\tilde{x}}_{12}|\bn{I}]=-\frac{1}{m_e}\bn{p}_1|\bn{I}\rangle=-\frac{1}{m_e}\bn{p}_1|-\bn{I}\rangle,
\end{align}
we arrive at
\begin{align}
    \hat{\tilde{x}}_{12}\bn{[12]}\left(\bn{[3-I]\aaaa{45}+\aaaa{35}[4-I]}\right)\bn{\ssaa{-I|3|5}}&=-\frac{1}{m_e}\bn{[12]}\left(\bn{\aaaa{45}\ssaa{3|1|-I}+\aaaa{35}\ssaa{4|1|-I}}\right)\bn{\ssaa{-I|3|5}}\nonumber\\
    &=\frac{1}{m_e}\bn{[12]}\left(\bn{\aaaa{45}\ssaa{3|1|-I}\ssaa{-I|4|5}-\aaaa{35}\ssaa{4|1|-I}\ssaa{-I|3|5}}\right).
    \label{line21}
\end{align}
Using generalized Schouten identities and equations of motion, we get
\begin{align}
    \bn{[12]}\bn{\aaaa{45}}\bn{\ssaa{3|1|-I}}\bn{\ssaa{-I|3|5}}&=\bn{\aaaa{45}}\bn{\ssaa{-I|3|5}}\left(\bn{[32]\aass{-I|1|1}+[31]\aass{-I|I+2|2}}\right)\nonumber\\
    &=m_e\bn{\aaaa{45}}\left(\bn{[23]\aaaa{1-I}+[13]\aaaa{2-I}}\right)\bn{\ssaa{-I|3|5}}\nonumber\\
    &=m_e\bn{\aaaa{45}}\left(\bn{[23]\aaaa{1|-I 4|5}}+\bn{[13]\aaaa{2|-I 4|5}}\right).
    \label{line22}
\end{align}
At this point, it is helpful to pause a bit to organize the terms we have. Noticing that the two terms on the right-hand side of Eq.~\eqref{line21} are related to each other by the antisymmetric exchange of $\bn{3}$ and $\bn{4}$, and the two terms on the right-hand side of Eq.~\eqref{line22} are related by the symmetric exchange of $\bn{1}$ and $\bn{2}$, we can write
\begin{align}
    \hat{\tilde{x}}_{12}\bn{[12]}\left(\bn{[3-I]\aaaa{45}+\aaaa{35}[4-I]}\right)\bn{\ssaa{-I|3|5}}=&\bn{\aaaa{45}}\bn{[23]\aaaa{1|-I 4|5}}\nonumber\\
    &+\bn{1\leftrightarrow 2-3\leftrightarrow 4-(1,3)\leftrightarrow(2,4)}.
\end{align}
We apply generalized Schouten identities one more time and use momentum conservation to get
\begin{align}
    \bn{\aaaa{45}}\bn{[23]\aaaa{1|-I 4|5}}&=\bn{[23]}\left(\bn{\aaaa{5|-I 4|5}\aaaa{41}-\aaaa{51}\aaaa{4|-I 4|5}}\right)\nonumber\\
    &=\bn{[23]}\left(-\bn{\aaaa{5|34+44|5}\aaaa{41}}+\bn{\aaaa{51}\aaaa{4|34+44+54|5}}\right)\nonumber\\
    &=\bn{[23]}\left(\bn{\aaaa{5|34|5}\aaaa{14}}-\bn{\aaaa{15}\aaaa{4|34|5}}-\left(2\hat{p}_4\cdot \hat{p}_5+m_\mu^2\right)\bn{\aaaa{15}\aaaa{45}}\right).
    \label{line23}
\end{align}
Using $\bn{\aaaa{4|34|5}}=2\hat{p}_3\cdot\hat{p}_4\bn{\aaaa{45}}+m_\mu\bn{\ssaa{4|3|5}}$ and $(\hat{p}_3+\hat{p}_4+\hat{p}_5)^2=2\hat{p}_3\cdot \hat{p}_4+2\hat{p}_3\cdot \hat{p}_5+2\hat{p}_4\cdot \hat{p}_5+2m_\mu^2=0$, we can rewrite Eq.~\eqref{line23} as
\begin{align}
    \bn{\aaaa{45}}\bn{[23]\aaaa{1|-I 4|5}}=\bn{[23]}\left[\bn{\aaaa{14}\aaaa{5|34|5}}+\left(m_\mu^2+2\hat{p}_3\cdot \hat{p}_5\right)\bn{\aaaa{15}\aaaa{45}}-m_\mu\bn{\ssaa{4|3|5}\aaaa{15}}\right].
\end{align}
Therefore, the second line of Eq.~\eqref{12pole} can be rewritten as
\begin{align}
    &-\tilde{e}^3\hat{\tilde{x}}_{12}\bn{[12]}\frac{\left(\bn{[3-I]\aaaa{45}+\aaaa{35}[4-I]}\right)\bn{\ssaa{-I|3|5}}}{\left(2\bn{p}_3\cdot \bn{p}_5\right)\left(2\bn{p}_4\cdot \bn{p}_5\right)}\nonumber\\
    &=\frac{-\tilde{e}^3}{\left(2\bn{p}_3\cdot \bn{p}_5\right)\left(2\bn{p}_4\cdot \bn{p}_5\right)}\left[\bn{\aaaa{14}\aaaa{5|34|5}}\bn{[23]}+\left(m_\mu^2+2\hat{p}_3\cdot \hat{p}_5\right)\bn{\aaaa{15}\aaaa{45}}\bn{[23]}-m_\mu\bn{\ssaa{4|3|5}\aaaa{15}}\bn{[23]}\right]\nonumber\\
    &\quad+\bn{1\leftrightarrow 2-3\leftrightarrow 4-(1,3)\leftrightarrow(2,4)}.
    \label{secondline}
\end{align}
Putting Eq.~\eqref{firstline} and Eq.~\eqref{secondline} together, the terms which explicitly depend on $m_\mu$ cancel out, leaving us with
\begin{align}
    \sum_{\lambda}\hat{A}_3(\mathbf{12}I^\lambda) \times \hat{A}_4(\mathbf{34}5^-{-I}^{-\lambda}) \Big|_{\hat{p}_{12}^2=0}=&\left[\frac{-\tilde{e}^3}{\left(2\bn{p}_3\cdot \bn{p}_5\right)\left(2\bn{p}_4\cdot \bn{p}_5\right)}\bn{\aaaa{5|34|5}\aaaa{14}[23]}+\frac{-\tilde{e}^3}{2\bn{p}_4\cdot \bn{p}_5}\bn{\aaaa{15}\aaaa{45}[23]}\right] \nonumber \\
    &+\bn{1\leftrightarrow 2-3\leftrightarrow 4-(1,3)\leftrightarrow(2,4)},
\end{align}
which is Eq.~\eqref{eq:s_12}.

\bibliographystyle{utphys} 
\bibliography{references}

\end{document}